\documentclass[lettersize,journal]{IEEEtran}

\usepackage{amsmath,amsfonts}
\usepackage{algorithmic}
\usepackage{algorithm}
\usepackage{array}
\usepackage[caption=false,font=normalsize,labelfont=sf,textfont=sf]{subfig}
\usepackage{textcomp}
\usepackage{stfloats}
\usepackage[caption=false]{subfig}
\usepackage{url}
\usepackage{verbatim}
\usepackage{graphicx}
\usepackage{multirow}
\usepackage{color}

\usepackage{etoolbox}
\newcommand{\PreserveBackslash}[1]{\let\temp=\\#1\let\\=\temp}
\newcolumntype{C}[1]{>{\PreserveBackslash\centering}p{#1}}
\newcolumntype{R}[1]{>{\PreserveBackslash\raggedleft}p{#1}}

\usepackage{cite}
\begin{document}

\title{LFIC-DRASC: Deep Light Field Image Compression Using Disentangled Representation and Asymmetrical Strip Convolution}

\author{Shiyu Feng, Yun Zhang~\IEEEmembership{Senior Member,~IEEE}, Linwei Zhu, and Sam Kwong~\IEEEmembership{Fellow,~IEEE}
\thanks{Shiyu Feng and Linwei Zhu are with the Shenzhen Institute of Advanced Technology, Chinese Academy of Sciences, Shenzhen 518055, China (e-mail: \{sy.feng1, lw.zhu\}@siat.ac.cn).}
\thanks{Yun Zhang is with the School of Electronics and Communication Engineering, Sun Yat-Sen University, Shenzhen, China (e-mail: zhangyun2@mail.sysu.edu.cn).}
\thanks{Sam Kwong is with the Department of Computing and Decision Sciences,Lingnan University, Hong Kong, China (e-mail: samkwong@ln.edu.hk).}
}


\maketitle

\begin{abstract}


Light-Field (LF) image is emerging 4D data of light rays that is capable of realistically presenting spatial and angular information of 3D scene. However, the large data volume of LF images becomes the most challenging issue in real-time processing, transmission, and storage. In this paper, we propose an end-to-end deep LF Image Compression method Using Disentangled Representation and Asymmetrical Strip Convolution (LFIC-DRASC) to improve coding efficiency. Firstly, we formulate the LF image compression problem as learning a disentangled LF representation network and an image encoding-decoding network. Secondly, we propose two novel feature extractors that leverage the structural prior of LF data by integrating features across different dimensions. Meanwhile, disentangled LF representation network is proposed to enhance the LF feature disentangling and decoupling. Thirdly, we propose the LFIC-DRASC for LF image compression, where two Asymmetrical Strip Convolution (ASC) operators, i.e. horizontal and vertical, are proposed to capture long-range correlation in LF feature space. These two ASC operators can be combined with the square convolution to further decouple LF features, which enhances the model ability in representing intricate spatial relationships. Experimental results demonstrate that the proposed LFIC-DRASC achieves an average of 20.5\% bit rate reductions comparing with the state-of-the-art methods.
\end{abstract}


\section{Introduction}
\IEEEPARstart{L}{ight}-Field (LF) imaging is an innovative and emerging technology offering users an immersive experience by capturing both spatial and angular information, thereby enabling the recording of Three-Dimensional (3D) geometry and light intensity efficiently. Different from panoramic video \cite{9833528}, multiview plus depth \cite{8416728}, and point cloud \cite{9502695} LF imaging facilitates interactive functionalities, including refocusing, perspective shifts, and the addition of augmented reality overlays. With its unique capabilities, LF imaging has become a promising technique for wide future media applications, such as lighting and rendering, refocusing cameras, synthetic aperture imaging, 3D displays and monitoring. However, the large size of LF data hinders its widespread application, which requires effective compression.

\begin{figure}
  \centering
  \subfloat[]{
    \includegraphics[width=0.18\textwidth]{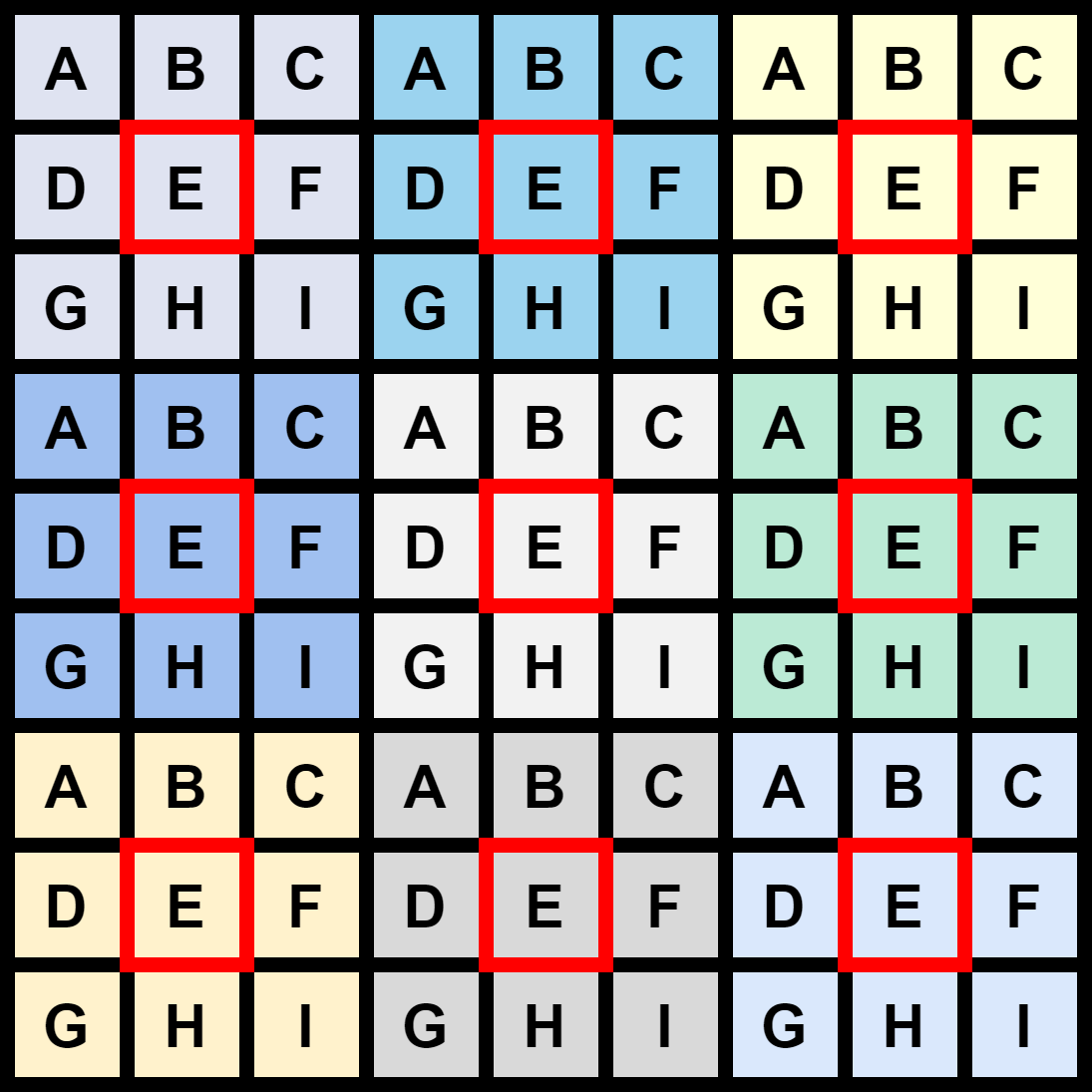}}
  \subfloat[]{
    \includegraphics[width=0.18\textwidth]{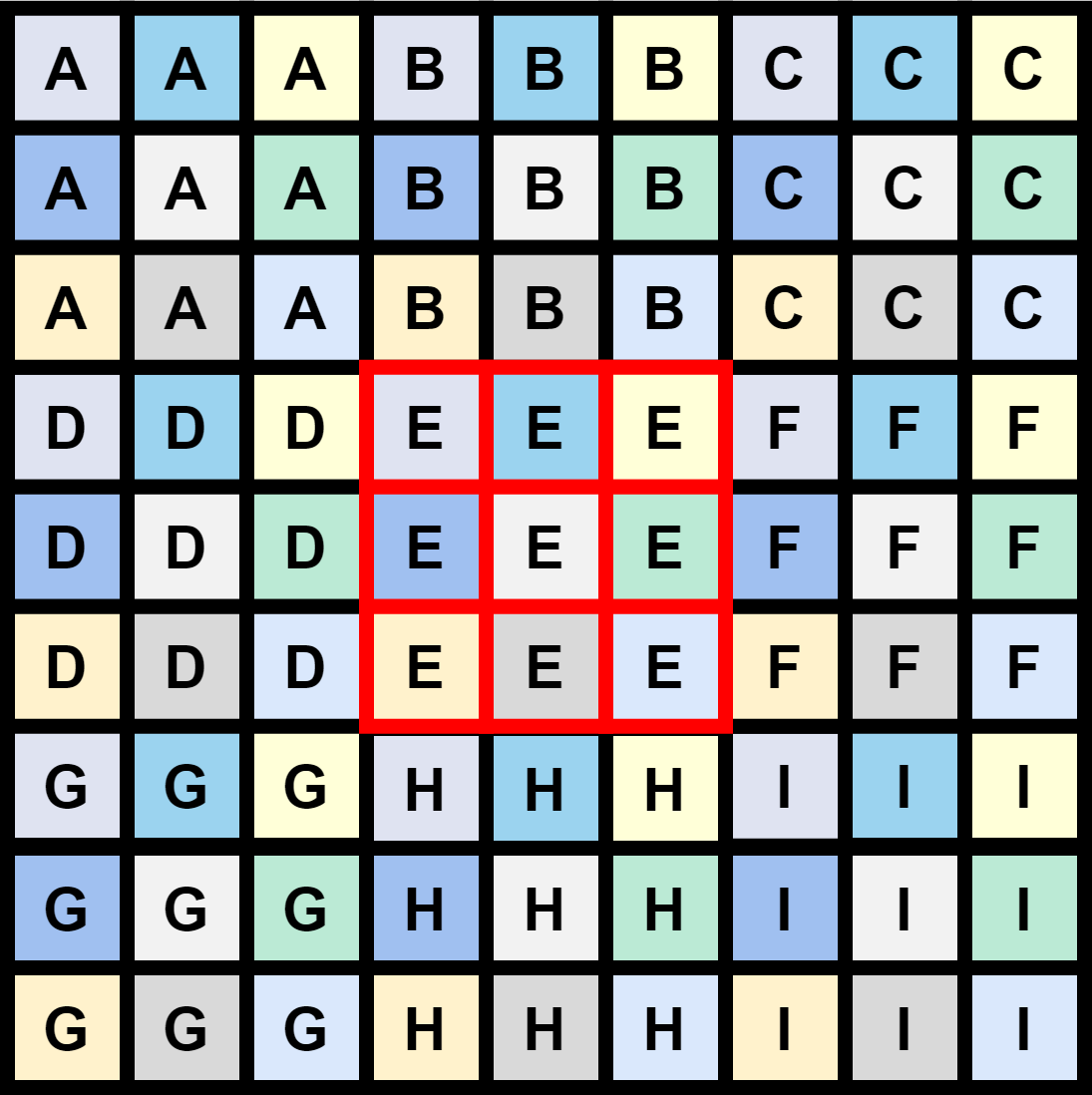}} \\
  \subfloat[]{
    \includegraphics[width=0.18\textwidth]{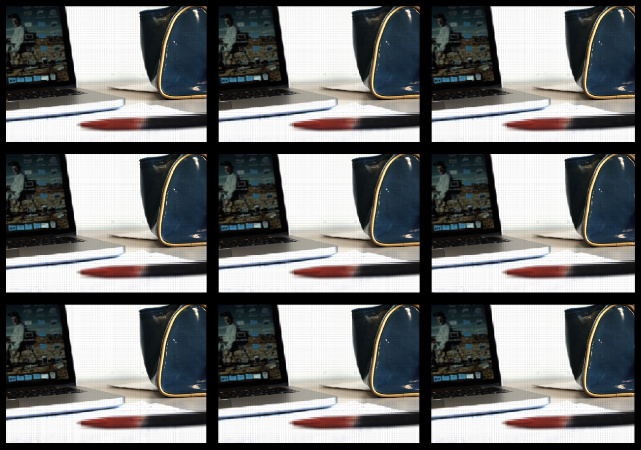}}
  \subfloat[]{
    \includegraphics[width=0.18\textwidth]{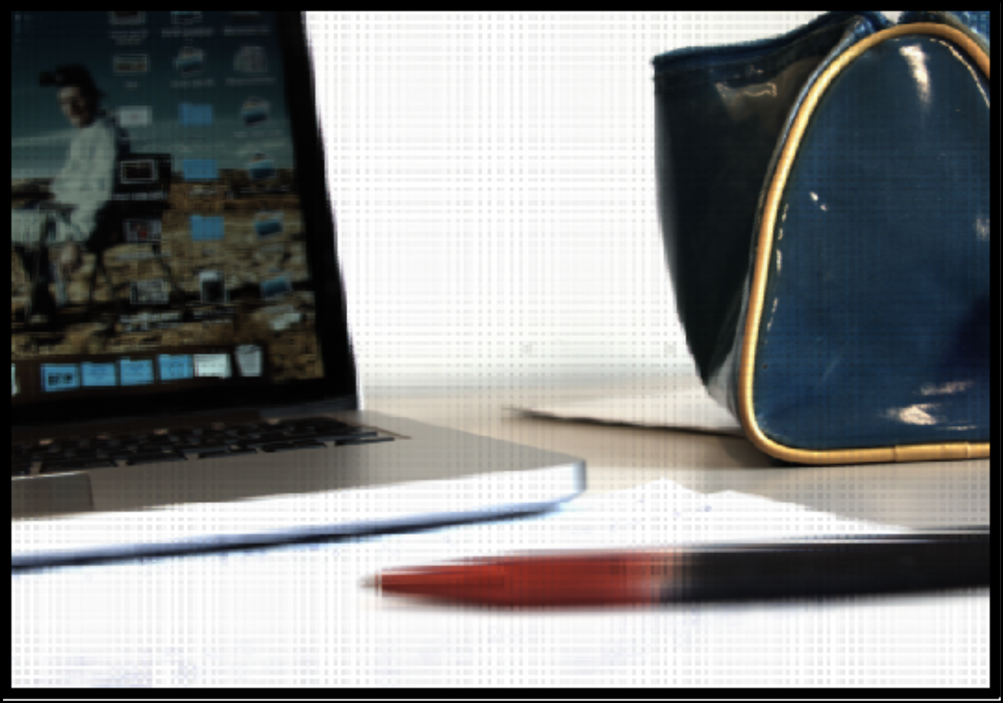}} \\
  \subfloat[]{
    \includegraphics[width=0.18\textwidth]{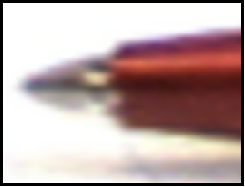}}
  \subfloat[]{
    \includegraphics[width=0.18\textwidth]{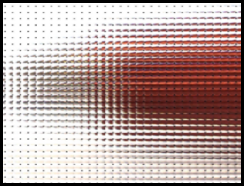}}
  \caption{SAI and MacPI representations of LF. (a) SAI representation. (b) MacPI representation. (c) SAI images. (d) MacPI image. (e)-(f) Enlarged SAI and MacPI images.}
  \label{fig:representation}
\end{figure}

LF represents 4D information of light rays and has various forms of representation. Lenslet image, the raw representation of LF, is captured by multiple tiny lenses, where each directly records the angular information of light rays. The relative positions of these tiny lenses capture spatial information. To represent the LF more effectively, multiple LF representations were developed.
The 4D LF data can be arranged into arrays as Sub-Aperture Images (SAIs), of which the spatial dimension is $H\times W$ and the angular dimension is $U\times V$, as shown in Fig. \ref{fig:representation}. Transform-based coding methods efficiently represent these SAIs due to strong spatial correlations among sub-views. The SAIs can be re-arranged into a Pseudo Video Sequence (PVS) and to be encoded by a video encoder. Epipolar Plane Images (EPIs) are formed by selecting rows or columns from the SAI, in which the slope of lines in EPIs representing the disparity is helpful in LF reconstruction and depth estimation. Macro-PIxel (MacPI) combines pixels with identical $h$ and $w$ positions across different $u$ and $v$ SAIs. Images and pixel correspondence between SAI and MacPI are shown in Fig. \ref{fig:representation}.

Many LF processing works were proposed by exploiting the MacPI representation as it can represent both spatial and angular information uniformly. Ahmed \textit{et al.} \cite{MACPIcvpr2023} converted SAIs into MacPIs and trained a deep neural network for LF super-resolution using the epipolar-spatial relationship. Wang \textit{et al.} \cite{DisentanglingPAMI} proposed to disentangle LFs by projecting MacPI into different subspaces. SAIs present spatial information clearly, but require convert them to PVS to extract angular information. Therefore, a decoupled selective matching network \cite{Liu2024LightFS} was proposed, which decoupled the LF into SAIs and EPIs to efficiently utilize global spatial and geometric correspondences. Both Lenset image and MacPI represent spatial and angular information within the LF, but MacPI representation is suitable for Convolutional Neural Network (CNN) processing. MacPI allows the LF to be further represented, decomposing complex spatial and angular relationships for efficient processing. Therefore, Liu \textit{et al.} \cite{Liu2023EfficientLF} proposed an efficient LF super-resolution by exploring multi-scale spatial-angular correlations and performing angular super-resolution on MacPI features.

To compress LF images more effectively, a number of LF coding algorithms have been proposed to exploit various LF representations. Lenset images consist of micro-images, allowing explicit extraction of variations in light. To compress the raw lenset LF images, inter-view prediction \cite{Mehajabin2022AnEP} and displacement based intra prediction \cite{6-18} were proposed based on High Efficiency Video Coding (HEVC) prediction tools, which exploited the inter-view and intra correlations in lenslet images. Furthermore, dictionary learning was utilized by representing adjacent microimages with sparse linear combinations \cite{6-20}. Liu \textit{et al.} \cite{Liu20245DEM} proposed a modeling-based compression approach for lenset images, which employed 5D epanechnikov kernel. However, lenset images lowered the spatial correlation among neighboring pixels, which reduces the coding efficiency. As SAIs present the LF spatial information well, graph lifting transform was proposed to improve SAI coding efficiency. Rizkallah \textit{et al.} \cite{Rizkallah2021RateDistortionOG} utilized local graph transform and partitioning to exploit the long-term signal correlation in LF coding. Zhang \textit{et al.} \cite{Zhang2023LightFC} proposed a dictionary learning and graph learning based LF image compression to explore structural redundancies among SAI views. Liu \textit{et al.} \cite{Liu2021ViewSL} proposed a Generative Adversarial Network (GAN) based reconstruction to reconstruct non-Key SAIs with key SAIs at client, which reduced the number of coded SAIs for high compression ratio. Amirpour \textit{et al.} \cite{Amirpour2022AdvancedSF} divided LF into sequential viewport layers and used previously encoded viewports for synthesis, which enhanced the viewport and quality scalability. Ahmad \textit{et al.} \cite{VVC21} categorized views into key and decimated views, where Shearlet-transform based prediction were used to predict decimated views. To explore the inter-view correlation among SAIs, SAIs are often organized in a pseudo time order and coded by video encoders with inter-prediction, such as HEVC \cite{VVC10} or Versatile Video Coding (VVC) \cite{VVC11}. Key views were encoded using multiview HEVC, while synthesized views and their residuals were encoded as a single PVS stream. Bakir \textit{et al.} \cite{VVC24} exploited temporal scalability of VVC to identify non-reference views who were reconstructed with GANs. A PVS-based Joint Exploration Model (JEM) for LF images captured by a Lytro lenset camera was proposed in \cite{K15}. Jia \emph{et al.} \cite{K31} proposed a LF image coding using GAN based view synthesis, which learned the angular and spatial context of LF image and synthesized intermediate SAIs. In \cite{K32}, SAIs was organized as PVS by minimizing the inter-correlations among views, which were encoded hierarchically with HEVC. Shi \textit{et al.} \cite{Shi2023LearningKN} proposed a deep representation network that reconstructed target SAIs from randomly initialized noise. This work employed modulator allocation and kernel tensor decomposition to enhance compaction. EPI presents the angular and spatial information of LF jointly, where line slopes present the geometry of objects. To compress the EPIs of LF effectively, Gao \textit{et al.} \cite{EPI-2019} proposed an EPI-based LF reconstruction framework to address angular restoration using a ``blur-restoration-deblur" approach. Sheng \textit{et al.} \cite{EPI-2018} handled occlusions by using multi-orientation EPIs for depth estimation. In \cite{EPI-code2}, key views of LF captured by parallel camera array were organized as EPIs and coded. Then, a Shearlet transform was applied iteratively to recover line slopes of sparse EPIs. However, EPI is challenging to discerning spatial relationships. Overall, every LF representation has its own advantages and disadvantages.

\begin{figure*}[!tb]
  \centering
  \subfloat[]{
    \includegraphics[width=0.7\textwidth]{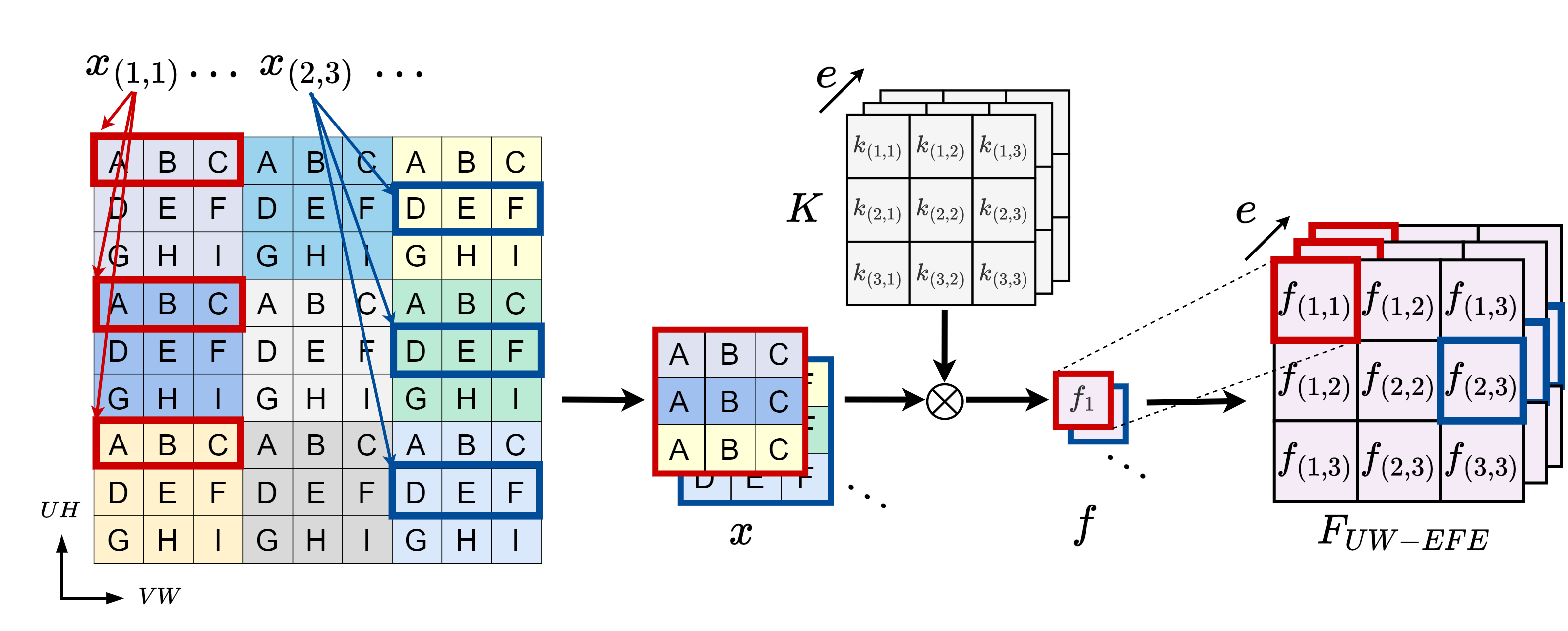}}
  \hspace{0.05cm}
  \subfloat[]{
    \includegraphics[width=0.7\textwidth]{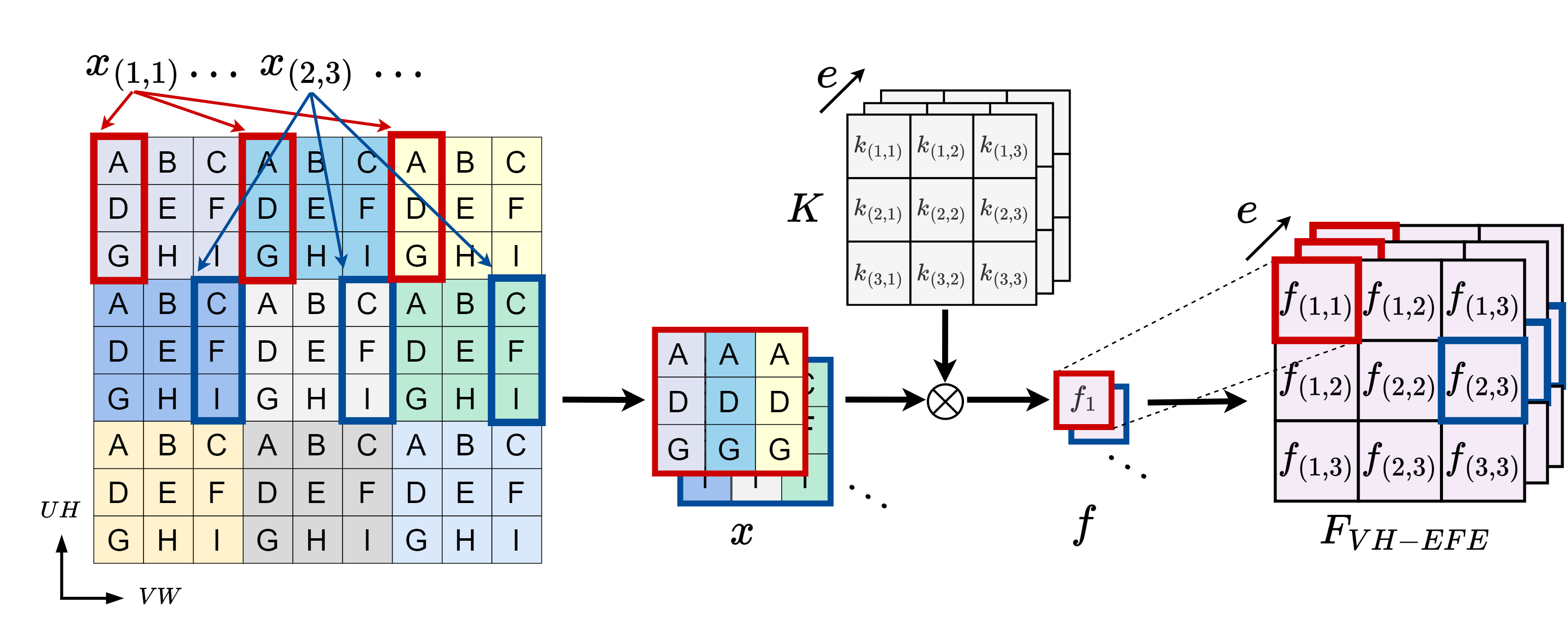}} 
      \hspace{0.05cm}
  \caption{Proposed EFE in U-W and V-H dimensions. (a) UW-EFE, (b) VH-EFE.}
  \label{fig-EFE}
\end{figure*}

End-to-end image compression employed a Variational AutoEncoder (VAE) \cite{VAE} to transform images into low-dimensional and compact representations, which achieved great success in 2D image compression\cite{9858899}. Cheng \textit{et al.} \cite{ChengCVPR2020} proposed a model based on residual blocks for image compression, while He \textit{et al.} \cite{Checkerboard} designed a checkerboard context model to facilitate parallel computing. However, LF images differs from traditional 2D images, and many network structures designed for 2D images are insufficient for high-dimensional LF representation. Firstly, the pixel correlation in LF images is not as smooth as in traditional 2D images \cite{Survey}, which is not conducive to existing CNN convolutions. Secondly, there are multiple kinds of LF representations presenting different types of angular and spatial information, having various spatial correlations and redundancies.
To compress the LF in end-to-end manner, Mohana \textit{et al.} \cite{PCS} proposed a LF compression framework based on SAI, where different SAIs were input to the network grouped in parallax for compression, and a new parallax-aware loss function was introduced. Tong \textit{et al.} \cite{SADN} built an end-to-end LF compression framework, where extract spatial-angular features were extracted by by a deep CNN and encoded by an entropy encoder. These research efforts achieved significant advancement in designing networks for LF image compression. However, they still could be further improved by further exploiting angular and spatial correlations and long-term dependencies in LF images.

In this work, we propose a deep LF Image Compression using Disentangled Representations and Asymmetrical Strip Convolution (LFIC-DRASC) to improve coding efficiency. The main contributions are
\begin{itemize}
  \item[$\bullet$]
 We formulate the LF image coding problem as learning a disentangled LF representation network and an image encoding-decoding network.
  \item[$\bullet$]
  We propose two novel feature extractors that leverage the structural prior of LF data by integrating features across dimensions and angles. Meanwhile, disentangled LF representation network is proposed to disentangle LF features and enhance the LF decoupling.
  \item[$\bullet$]
 We propose the LFIC-DRASC network for LF image compression, where two Asymmetrical Strip Convolution (ASC) operators, i.e. horizontal and vertical ones, are proposed to capture long-range correlation in LF feature space. The two ASC operators can be combined with the conventional square convolution kernel to further decouple the LF features, which enhances the model ability in representing intricate spatial relationships.
\end{itemize}
The rest of this paper is organized as follows. Section II presents problem formulation of LF image compression. Section III introduces the LF feature disentangling network. Section IV presents the framework of the LFIC-DRASC and ASC. Section V presents the experimental results and analyses. Section VI draws the conclusions.

\section{Problem Formulation}
LF images are 4D information of light rays $\mathcal{L}(u, v, h, w) \in\mathbb{R}^{U \times V \times H \times W}$, where ($u$, $v$) and ($h$, $w$) represent the angular and spatial information of light rays, respectively. Since LFs are high-dimensional data with various representations, we aim to extract appropriate features for compression. The LF representation and compression problem is formulated to minimize the distortion between the original and reconstructed LF images subject to bit rate constraints, which is
\begin{equation}
\begin{cases}
\min {D}({\mathbf{L},\hat{\mathbf{L}})}, s.t. R(\mathbb{M}_E(\mathbf{F},Q))\leq R_T \\
\mathbf{F}=\mathbb{R}(\mathbf{L}) \\
\hat{\mathbf{L}}=\mathbb{M}_D(\mathbb{M}_E(\mathbf{F},Q))
\end{cases},
\end{equation}
where $\mathbb{R}()$ is a LF representation module to project raw $\mathbf{L}$ to feature set $\mathbf{F}$, $\mathbb{M}_E$ and $\mathbb{M}_D$ are LF image encoding and decoding modules, where $Q$ is the quantization factor, ${D}$ is the distortion metric for LF, $R()$ is the function of bit rate,  $R_T$ is the target bit rate. The conditional constrained problem can be converted to the optimization problem as
\begin{equation}
\{\mathbb{M}_E,\mathbb{M}_D,\mathbb{R}\}^* \\
 = \arg\min \limits_{\{\mathbb{R},\mathbb{M}_E,\mathbb{M}_D\}} {D}({\mathbf{L},\hat{\mathbf{L}})}+\lambda R(\mathbb{M}_E(\mathbf{F},Q)),
\end{equation}
where $\lambda$ is the Lagrange multiplier. The objective is to find the optimal LF representation $\mathbb{R}$, and LF compression networks $\{\mathbb{M}_E, \mathbb{M}_D\}$ at the minimum rate-distortion cost. Sections III and IV present the designs of $\mathbb{R}$ and $\{\mathbb{M}_E, \mathbb{M}_D\}$, respectively.

\begin{figure*}[!tb]
  \centering
  \includegraphics[width=0.75\textwidth]{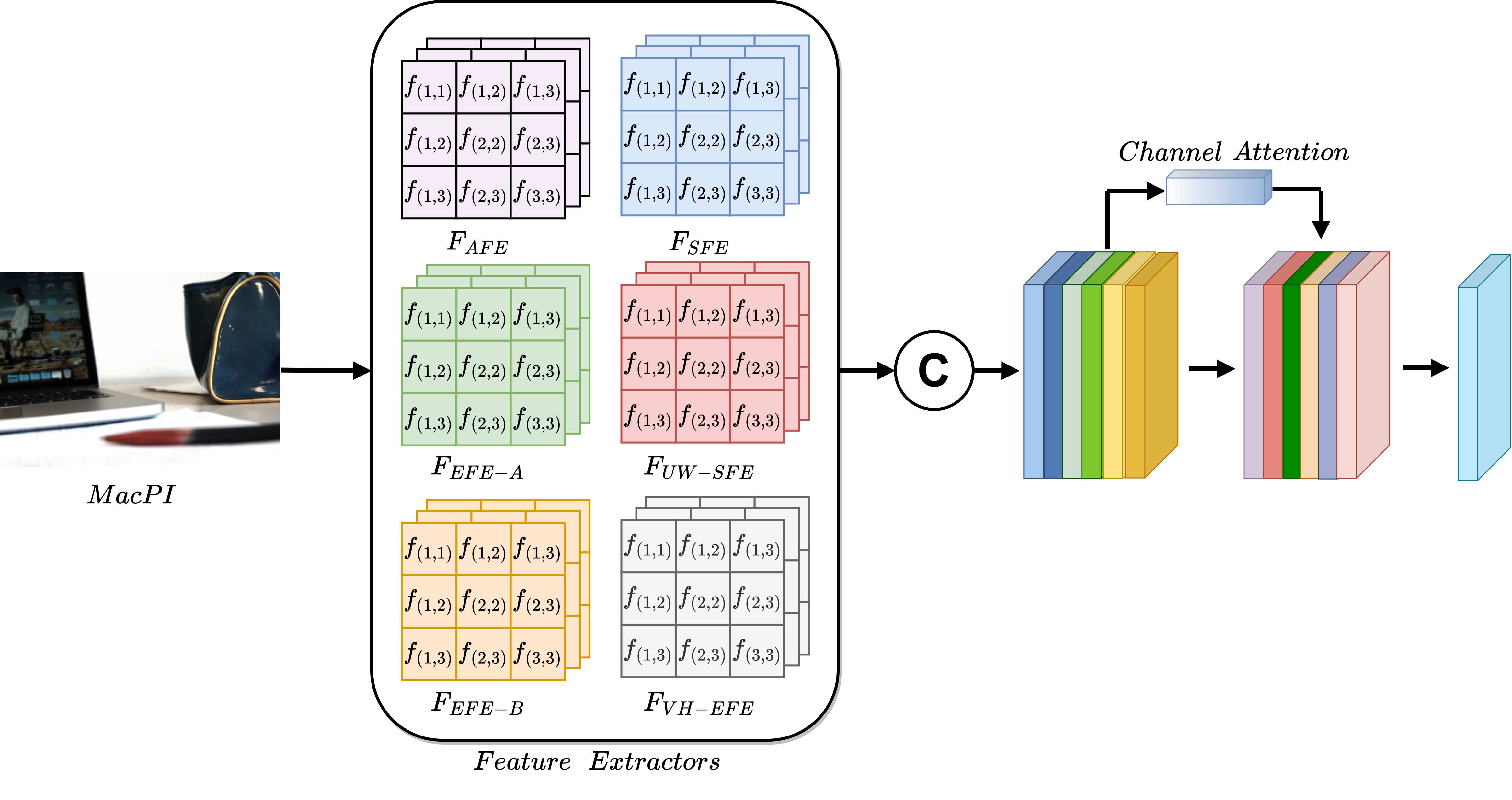}
  \caption{Architecture of the proposed FDM, where $\mathbf{F}_{AFE}$, $\mathbf{F}_{SFE}$, $\mathbf{F}_{EFE-A}$, and $\mathbf{F}_{EFE-B}$ are from \cite{wang2020spatialangular}, $\mathbf{F}_{VH-EFE}$ and $\mathbf{F}_{UW-EFE}$ are proposed.}
  \label{fig:disentangling block}
  \end{figure*}

\section{Learning Disentangled Representation for LF Image Coding}
\subsection{Proposed LF Feature Extractors}
To extract effective LF features, the Spatial Feature Extractor (SFE) and Angular Feature Extractor (AFE) operate within the (H-W) and (U-V) subspaces, respectively, while the EPI Feature Extractor (EFE)-A and EFE-B operate within the (U-H) and (V-W) subspaces, respectively \cite{wang2020spatialangular}. However, the spatial dimension combinations (V-H) and (U-W) in 4D data have been overlooked. To address this problem, we propose a more comprehensive disentangled LF feature representation for $\mathbb{R}$, which can be formulated as
\begin{equation}
\begin{cases}
  \mathbf{F}(i,j,a,b) = \mathbb{R}(\mathbf{L}(u,v,h,w)) \\
  \mathbb{R}(x_{a,b}) = \sum\limits_{m,n,e}K_{m,n,e} x_{a+m,b+n} + b_{a,b}
\end{cases},
\end{equation}
where $\mathbf{L}(u,v,h,w)$ is the input of LF image, $\mathbf{F}(i,j,a,b)$ is the output feature map of LF, $K_{m,n,e}$ is filter parameter, $b_{a,b}$ is bias at position $(a,b)$. It indicates that the convolution $K_{m,n,e}$ is multiplied by elements of $L$, and shifted by $m-1$ and $n-1$ around the point $(a,b)$.

We propose two new operators, \emph{i.e.}, UW-EFE and VH-EFE, to capture relationships within the U-H and V-W spaces. Specifically, the proposed feature extractors UW-EFE and VH-EFE also convolve pixels from the EPI, but they work in the V-H and U-W subspaces, respectively. For the UW-EFE and VH-EFE in the H-V subspace, as shown in Fig. \ref{fig-EFE}, the red and blue lines indicate the convolutional ranges of UW-EFE and VH-EFE when the convolutional kernels move to different positions. The UW-EFE and VH-EFE feature extractors are
\begin{equation}
\begin{cases}
  \mathbf{F}_{UW-EFE}(u,h,a,b)= \mathbb{R}(\mathbf{L}(u, a+m-1, h, b+n-1))\\
  \mathbf{F}_{VH-EFE}(v,w,a,b)= \mathbb{R}(\mathbf{L}(v, a+m-1, w, b+n-1))
\end{cases},
\end{equation}
where $m$ and $n$ are offsets. The UW-EFE and VH-EFE feature extractors are designed to complete the projection of the LF by filling in the missing subspaces. However, the UW-EFE feature extractor is not composed of a simple single convolution kernel. Instead, we employ a two-layer convolution approach for the UW-EFE feature extractor. In the first layer, the convolution kernel size is $1\times A$, with a vertical stride of 1 and a horizontal stride of A, producing an output feature space of $AH\times W$. The second layer convolves over the feature space output from the first layer using a kernel size of $A\times 1$, with both vertical and horizontal strides of 1. This method effectively splits the convolution of UW-EFE into these two distinct kernels. Altogether, six operators now disentangle the 4D LF space (U$\times$V$\times$H$\times$W) into six subspaces: U-W, U-H, U-V, V-H, V-W, and H-W, expanding beyond the original four subspaces. This comprehensive set of solutions allows for the complete disentanglement of the 4D LF into 2D subspaces. These feature extractors collaborate to create a vast receptive domain, integrate information across different spaces, capture any subtle changes in the LF data, and enable the network to model the LF in multiple dimensions.

\begin{figure*}[!tb]
  \centering
  \includegraphics[width=0.9\textwidth]{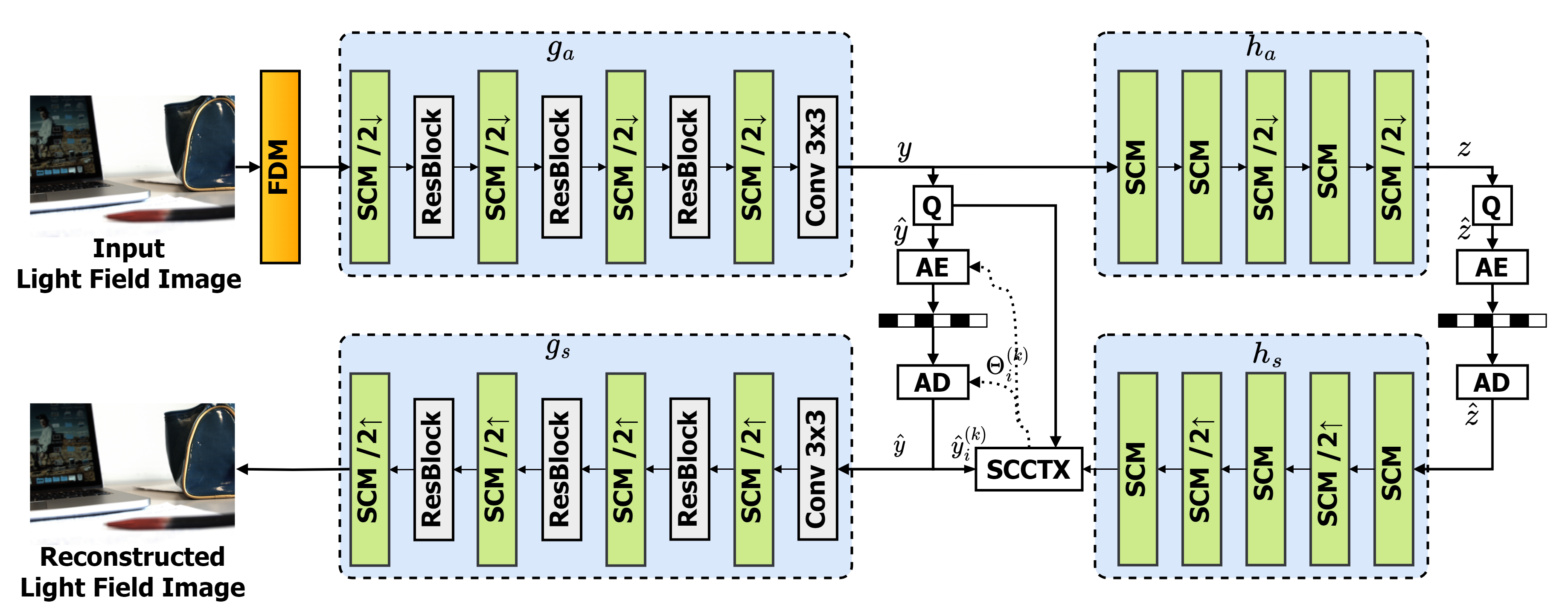}
  \caption{The proposed framework of LFIC-DRASC.}
  \label{fig:network map}
\end{figure*}

\subsection{Feature Disentangling Module (FDM) of LF}

The FDM aims to extract representative LF features for effective compression, which involves a series of feature disentangling modules on different LF subspaces. Firstly, the input LF array is transformed into the MacPI format and split into small patches. Then, these patches are processed using various operators to extract features from different subspaces. Finally, the extracted features from these subspaces provide a high-level representation for effective LF compression.

We first introduce the UW-EFE and VH-EFE operators in the FDM to fill in the neglected subspaces (V-H and U-W) in LF data, thereby achieving comprehensive disentanglement of the LF. Specifically, UW-EFE and VH-EFE convolve pixels on the EPI in the V-H and U-W subspaces, respectively. This approach allows us to capture more information, making feature extraction more comprehensive and accurate.
\begin{figure}[!tb]
  \centering
  \includegraphics[width=0.24\textwidth]{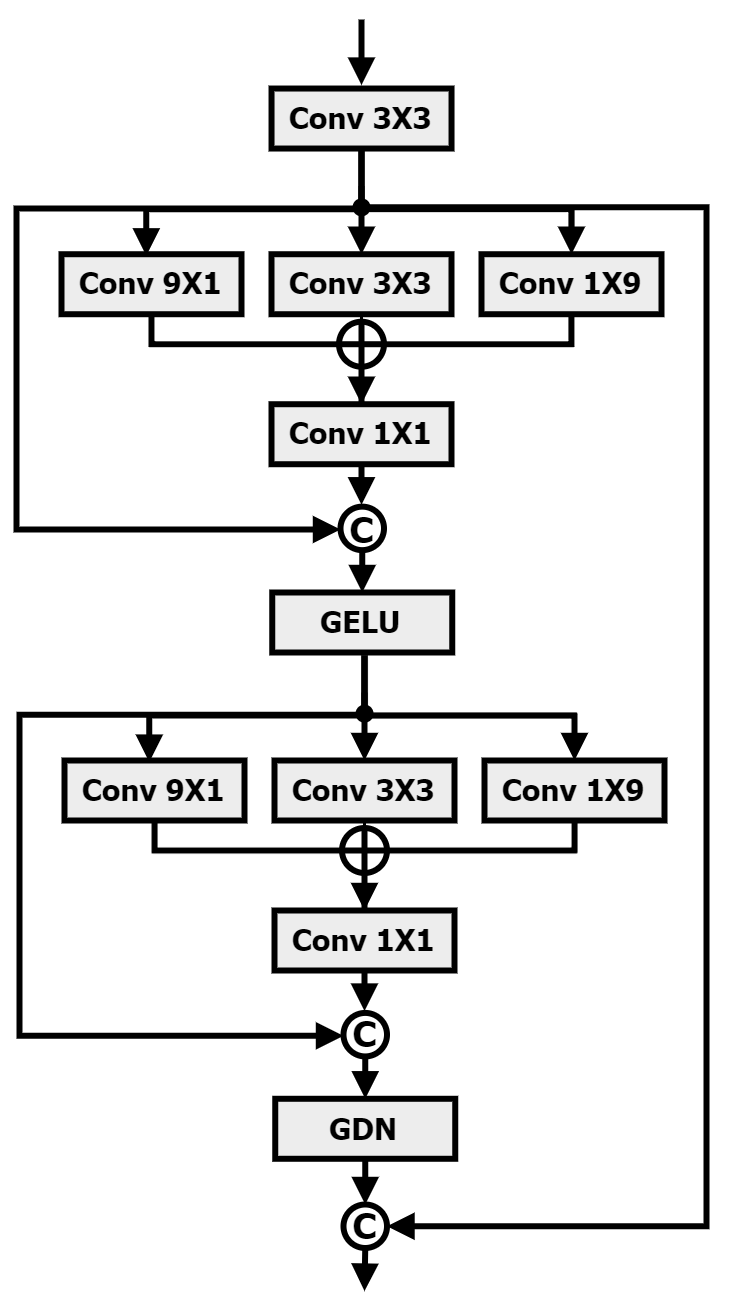}
  \caption{Structure of the SCM using horizontal and vertical ASCs.}
  \label{fig:network map2}
\end{figure}
Then, we fuse all the extracted features. By concatenating the features extracted by SFE, AFE, EFE-A, EFE-B, UW-EFE, and VH-EFE, we obtain a high-dimensional feature representation that covers the spatial domain (U-V), angular domain (H-W), and epipolar plane domain (U-H, U-W, and V-H, V-W). Since it is possible to consider that each subspace can be regarded as EPIs constructed based on SAIs with different stacking orders, we divided the subspaces reflecting the same spatial stacking order into one group, \emph{i.e.}, subspace U-W and subspace U-H into one group, while subspace V-W and  subspace V-H into another group. This fusion method ensures that information from different subspaces can complement each other, enhancing the LF feature representation.

To further enhance the effectiveness of feature extraction, we introduce a channel attention network, which adaptively assigns different weights to features based on their importance. By combining multiple attention-weighted features, the FDM ensures that the most significant features are emphasized, thereby improving the performance of LF data processing. 
FDM is mathematically presented as
\begin{equation}
  \begin{aligned}
  \mathbf{F} &= f_A\big(f_{C}(\mathbf{F}_{SFE}, \mathbf{F}_{AFE}, \mathbf{F}_{EFE-A}, \mathbf{F}_{EFE-B}, \\ &\quad \mathbf{F}_{UW-EFE}, \mathbf{F}_{VH-EFE})\big)
\end{aligned},
\end{equation}
where $f_A()$ and $f_C()$ are channel attention network and concatenate operation.


\section{The Proposed Deep LFIC-DRASC}

\subsection{Framework of the Deep LFIC-DRASC}

\begin{figure*}[t]
  \centering
  \subfloat[]{\includegraphics[height=0.24\textwidth]{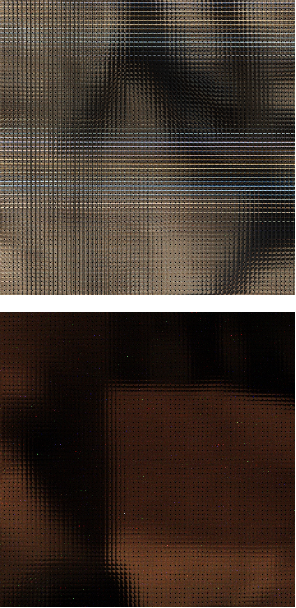}}
  \hfill
  \subfloat[]{\includegraphics[height=0.24\textwidth]{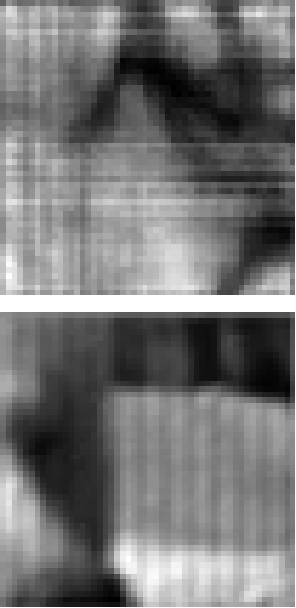}}
  \hfill
  \subfloat[]{\includegraphics[height=0.24\textwidth]{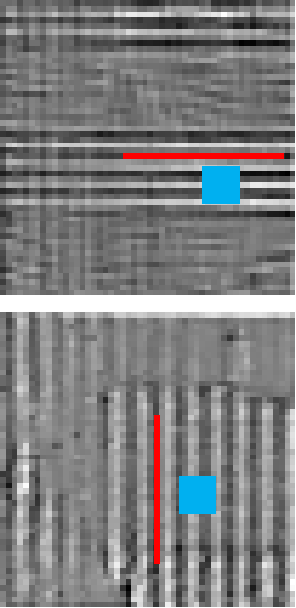}}
  \hfill
  \subfloat[]{\includegraphics[height=0.24\textwidth]{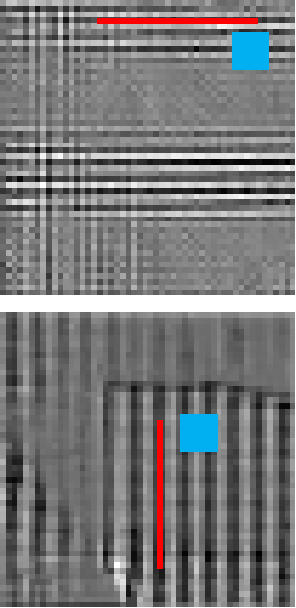}}
  \hfill
  \subfloat[]{\includegraphics[height=0.24\textwidth]{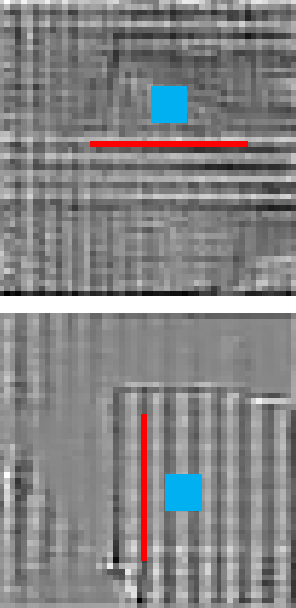}}
  \hfill
  \subfloat[]{\includegraphics[height=0.24\textwidth]{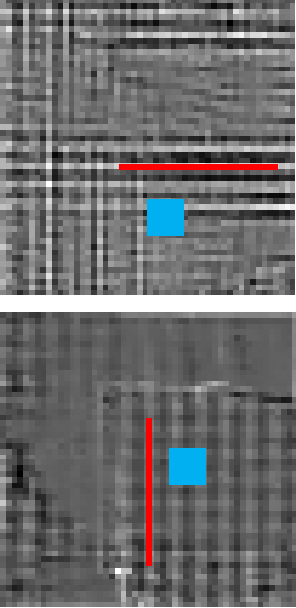}}
  \hfill
  \subfloat[]{\includegraphics[height=0.24\textwidth]{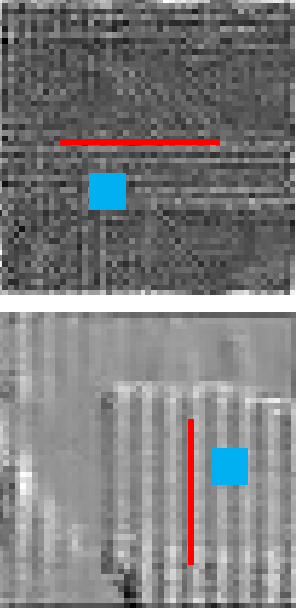}}
  \hfill
  \subfloat[]{\includegraphics[height=0.24\textwidth]{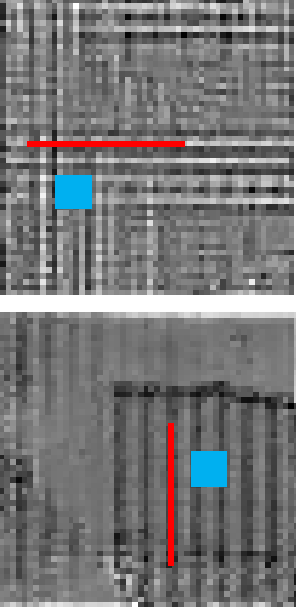}}
  \caption{Examples of strip feature maps. (a) The original LF MacPI patch. (b)-(h) Typical strip feature maps in the LF MacPI. }
  \label{fig:sc}
\end{figure*}

Fig. \ref{fig:network map} shows the overall structure of the end-to-end deep LFIC-DRASC compression network, which consists of FDM for LF feature disentangling and VAE based image compression network using ASC. ResBlock denotes residual bottleneck blocks and SCM is a subnetwork using the ASC, as shown in \ref{fig:network map2}. AE and AD are the arithmetic encoder and decoder, respectively. $Q$ denotes the quantization step. The encoder $\mathbb{M}_E$ and decoder $\mathbb{M}_D$
of the framework can be represented as
\begin{equation}
  \begin{cases}
    \boldsymbol{y}=g_{a,SCM}(\mathbb{R}(\boldsymbol{L}) ; \phi_{g}) \\
    \hat{\boldsymbol{y}}=q(\boldsymbol{y},Q) \\
    \hat{\boldsymbol{L}}=g_{s,SCM}(\hat{\boldsymbol{y}} ; \theta_{g})
\end{cases},
\end{equation}
where $\boldsymbol{L}$, $\boldsymbol{y}$, $\hat{\boldsymbol{y}}$, $g_{a,SCM}$ , $q$ and $g_{s,SCM}$ represent input source LF, the latent representation before quantization, the encoded representation, the main encoder, the quantization function, and the main decoder, respectively. $\phi_{g}$ and $\theta_{g}$ are the parameters of the encoder and decoder, respectively.
The hyperprior encoder and decoder are represented as

\begin{equation}
  \begin{cases}
    \boldsymbol{z}=h_{a,SCM}(\boldsymbol{y},\phi_h)\\
    \boldsymbol{\hat{z}}=q(\boldsymbol{z},Q) \\
    p_{\hat{\boldsymbol{y}} \mid \hat{\boldsymbol{z}}}(\hat{\boldsymbol{y}} \mid \hat{\boldsymbol{z}}) \leftarrow h_{s,SCM}\left(\hat{\boldsymbol{z}} ; \theta_{h}\right)
\end{cases},
\end{equation}
where $\boldsymbol{z}$ is the side information extracted from $\boldsymbol{y}$, $\boldsymbol{\hat{z}}$ represents its quantized version, $h_{a,SCM}$ and $h_{s,SCM}$ are the encoder and decoder of the hyper-prior part, respectively. $p_{\hat{\boldsymbol{y}} \mid \hat{\boldsymbol{z}}}(\hat{\boldsymbol{y}} \mid \hat{\boldsymbol{z}})$ is the estimated distribution conditioned on $\boldsymbol{\hat{z}}$.

The LF data is first represented as MacPI and input to the FDM to be represented in 2D subspaces. These 2D features are then input into a compression model based on VAE that utilizes a super-prior architecture and a context model for entropy estimation. The process starts with the FDM disentangling the input MacPI into a lower-dimensional form. This representation is then encoded into a latent space, followed by quantization to form a discrete latent representation. The decoder reconstructs the MacPI from this quantized latent representation.
We keep the traditional square kernel convolution while adding different strip convolutions to better fit the feature representation of the data in the LF. Specifically, we put four SCMs in each of the encoder and decoder, and replace all the traditional Resnet modules with SCM modules in the hyper-prior codec block. In order to efficiently predict the probability distribution of potential factors and reduce the bit rate effectively, we use the Space-Channel ConTeXt (SCCTX) based entropy model\cite{ELIC}.

By transforming LF images into lower-dimensional spaces and processing them through attention mechanisms, we can focus on the most relevant features, making the data easier to process and compress. The disentangled and reconstructed features are easier for the VAE to model, leading to more efficient learning and better reconstruction. This is because the VAE can concentrate on capturing the essential characteristics of the data without being overwhelmed by noise and redundancy. Potential variants to this structure could include different types of feature disentangling modules or alternative convolutional architectures within the VAE framework. For instance, one could explore the use of attention mechanisms within transformers or graph-based convolutions to further enhance the feature representation and compression. By integrating these advanced techniques, the network could achieve even higher compression rates and better reconstruction quality, paving the way for a more efficient LF image compression.

\begin{table*}
  \centering
  \caption{BD-Rate and BD-PSNR of LFIC-DRASC Comparing with Other LF Compression Schemes.}
  \label{table:Quantitative}

  \begin{tabular}{l|cc|cc|cc|cc|cc}
    \hline
    \multirow{3}{*}{\text { Images }} &\multicolumn{2}{c|}{{ Pro. vs. GCC}} & \multicolumn{2}{c|}{{ Pro. vs. Cheng's}} & \multicolumn{2}{c|}{{ Pro. vs. SOP-HEVC}} & \multicolumn{2}{c|}{{ Pro. vs. SOP-VVC}} & \multicolumn{2}{c}{{ Pro. vs. SADN}} \\
    \cline { 2 - 11 }

    & \multicolumn{1}{c}{BD-PSNR} & \multicolumn{1}{c|}{BD-BR} & \multicolumn{1}{c}{BD-PSNR} & \multicolumn{1}{c|}{BD-BR} & \multicolumn{1}{c}{BD-PSNR} & \multicolumn{1}{c|}{BD-BR} & \multicolumn{1}{c}{BD-PSNR} & \multicolumn{1}{c|}{BD-BR} & \multicolumn{1}{c}{BD-PSNR} & \multicolumn{1}{c}{BD-BR} \\
    & (dB) & (\%) & (dB) & (\%) & (dB) & (\%) &  (dB) & (\%) &  (dB) & (\%)\\
    \hline
    \text { Bikes }& +2.20 & -65.2 & +3.25 & -70.6 & +4.11 & -83.7 & +3.47 & -78.1 & +0.37 & -8.4 \\
    \text { Danger De Mort }& +1.73 & -53.5 & +1.83 & -53.1 & +3.86 & -72.1 & +2.61 & -62.3 & -0.44 & +31.5 \\
    \text { Flowers }& +3.21 & -68.9 & +3.27 & -74.4 & +3.00 & -83.2 & +2.36 & -79.2 & +0.76 & -39.3 \\
    \text { Stone Pillars Outside}& +3.28 & -55.8 & +2.47 & -63.5 & +2.94 & -82.5 & +2.21 & -76.3 & +1.39 & -47.7 \\
    \text { Vespa }& +1.18 & -41.0 & +3.66 & -58.6 & +3.21 & -76.7 & +1.99 & -66.7 & +1.26 & -41.1 \\
    \text { Ankylosaurus\&Diplodocus }& -1.25 & +64.2 & +5.26 & -86.1 & +0.17 & -6.7 & -0.53 & +54.2 & +0.97 & -25.1 \\
    \text { Desktop}& +0.03 & +1.5 & +3.13 & -68.7 & +1.97 & -61.5 & +0.63 & -41.3 & +1.09 & -47.1 \\
    \text { Magnets }& -0.96 & +181.0 & +3.82 & -95.4 & +0.42 & -16.7 & -0.14 & +15.5 & +0.42 & -14.6 \\
    \text { Fountain\&Vincent }& +2.56 & -65.8 & +2.61 & -65.2 & +2.40 & -79.6 & +1.26 & -68.4 & +1.06 & -34.2 \\
    \text { Friends }& +0.18 & -5.3 & +3.43 & -65.6 & +2.61 & -66.3 & +1.55 & -55.8 & +0.40 & -16.7 \\
    \text { Color Chart }& -0.01 & -3.9 & +5.14 & -84.9 & +0.35 & -25.6 & -0.69 & +59.5 & -0.56 & +24.4 \\
    \text { ISO Chart }& +2.30 & -69.5 & +2.36 & -70.2 & +1.73 & -63.0 & +0.52 & -26.8 & +1.20 & -28.0 \\
    \hline {\text { Average }} & \textbf{+1.20} & \textbf{-18.5} & \textbf{+3.35} & \textbf{-71.4} & \textbf{+2.23} & \textbf{-59.8}& \textbf{+1.27} & \textbf{-35.5}& \textbf{+0.66} & \textbf{-20.5}  \\
    \hline
  \end{tabular}
\end{table*}

\subsection{The Proposed SCM for LF Compression}
In the conventional end-to-end image compression process, square convolution kernel is typically employed to extract various features within a rectangular region, which is effective for conventional natural images. However, due to the spatial and angular representations of LF images \cite{SADN} and \cite{DisentanglingPAMI}, the square convolution may not be always effective. We extracted and illustrated the different layers of feature maps while encoding the LF images, as shown in Fig. \ref{fig:sc}. Due to the lens imaging properties of LF images, we can observed that there exist numerous striped texture structures along the horizontal and vertical axes. Compared to the traditional square kernel convolution (depicted by blue rectangles), the narrow shape of stripe convolution (illustrated by red rectangles) is more effective in capturing the abundant repetitive strip-like features presented in LF images.

Conventional square convolution are inadequate for capturing the strip-like features of LF images. Furthermore, it was found that increasing the receptive field of the backbone network enhances scene parsing capabilities \cite{Chen2018EncoderDecoderWA, Fu2018DualAN}. Strip convolution, with its long and narrow kernel shapes, can more effectively establish long-term dependencies between discretely distributed regions and encode striped areas, offering a significant advantage over the square convolution \cite{stripConv}. It also focuses on capturing local details due to its narrow kernel shape along the other dimension. Therefore, we propose the horizontal and vertical ASC operators, which have wider receptive fields and are able to further disentangle the 2D LF subspace along the horizontal and vertical axes.

\begin{figure*}[!tb]
  \centering
  \includegraphics[width=0.7\textwidth]{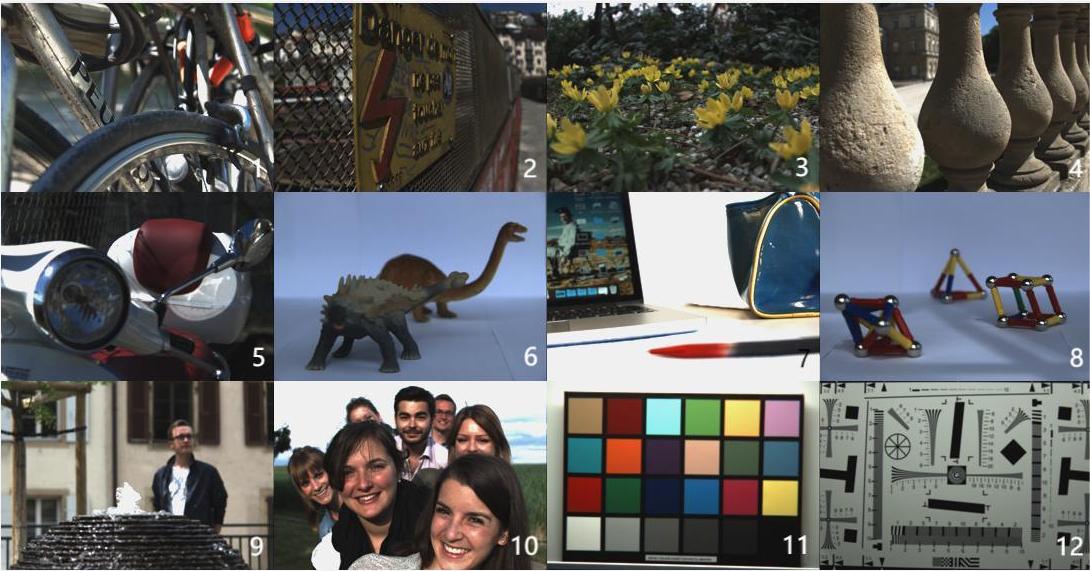}
  \caption{Thumbnails of the dataset for LF image compression \cite{2016ICME}. (1) Bikes. (2) Danger De Mort. (3) Flowers. (4) Stone Pillars Outside. (5) Vespa. (6) Ankylosaurus\&Diplodocus. (7) Desktop. (8) Magnets. (9) Fountain\&Vincent. (10) Friends. (11) Color Chart. (12) ISO Chart.}
  \label{fig:all}
\end{figure*}

Fig. \ref{fig:network map2} shows the structure of the proposed SCM. We replace the single square kernel with three dynamically sized convolution kernels based on the side length $D$. Specifically, we employ Conv $D\times D$, Conv $D^2 \times 1$, and Conv$1\times D^2$ to adapt to different feature dimensions. This configuration allows each kernel to capture features at different scales and orientations while maintaining an equivalent receptive field. The extracted features are then processed through a Conv$1 \times 1$ to fuse the information from different convolution kernels while maintaining the same number of channels. Within the SCM, the first strip convolution layer utilizes Gaussian Error Linear Unit (GELU) to perform a simple linear transformation of the features. Then, a Generalized Divisive Normalization (GDN) layer is used after the second strip convolution layer. To reduce the feature dimensions and the computational complexity in the end-to-end image compression, we integrate downsampling and inverse convolution within the SCM, along with upsampling to restore the original data dimensions. In this work, $D$ was set to 3, and the convolution kernels in the $k$th layer of the SCM are Conv$9 \times 1$, Conv$1 \times 9$, Conv$3 \times 3$ and Conv$1\times 1$, denoted as $C_{1}^{k}$, $C_{2}^{k}$, $C_{3}^{k}$ and $C_{0}^{k}$, respectively. The feature extraction in the ASC layer is presented as
\begin{equation}
  f_k(\mathbf{x}) = C_0^k ( \sum_{i=1}^3 C_i^k (\mathbf{x})) + \mathbf{x},
\end{equation}
where $C_i^k (\mathbf{x})$ represents the convolution operation of the $i$th kernel in the $k$th layer applied to the input $\mathbf{x}$. The overall SCM process is presented as
\begin{equation}
    \mathbf{F}_{\text{SCM}}(\mathbf{x}) = \operatorname{GDN}\left(f_2\left(\operatorname{GELU}\left(f_1\left(\mathbf{x}_{\text{up/down}}\right)\right)\right)\right)  + \mathbf{x}_{\text{up/down}},
\end{equation}
where $\mathbf{F}_{\text{SCM}}$ denotes the final output of the SCM, reflecting the latent representation after feature extraction and normalization, $\mathbf{{x}_\text{up/down}}$ represents the input of the SCM after it has undergone either upsampling or downsampling. The functions $f_1$ and $f_2$
represent the operations within the first and second strip convolution blocks respectively, which process the latent representation sequentially.

\begin{figure*}[!htbp]
  \centering
  \captionsetup[subfloat]{font=footnotesize}
  \subfloat[Bikes]{\includegraphics[width=0.25\linewidth]{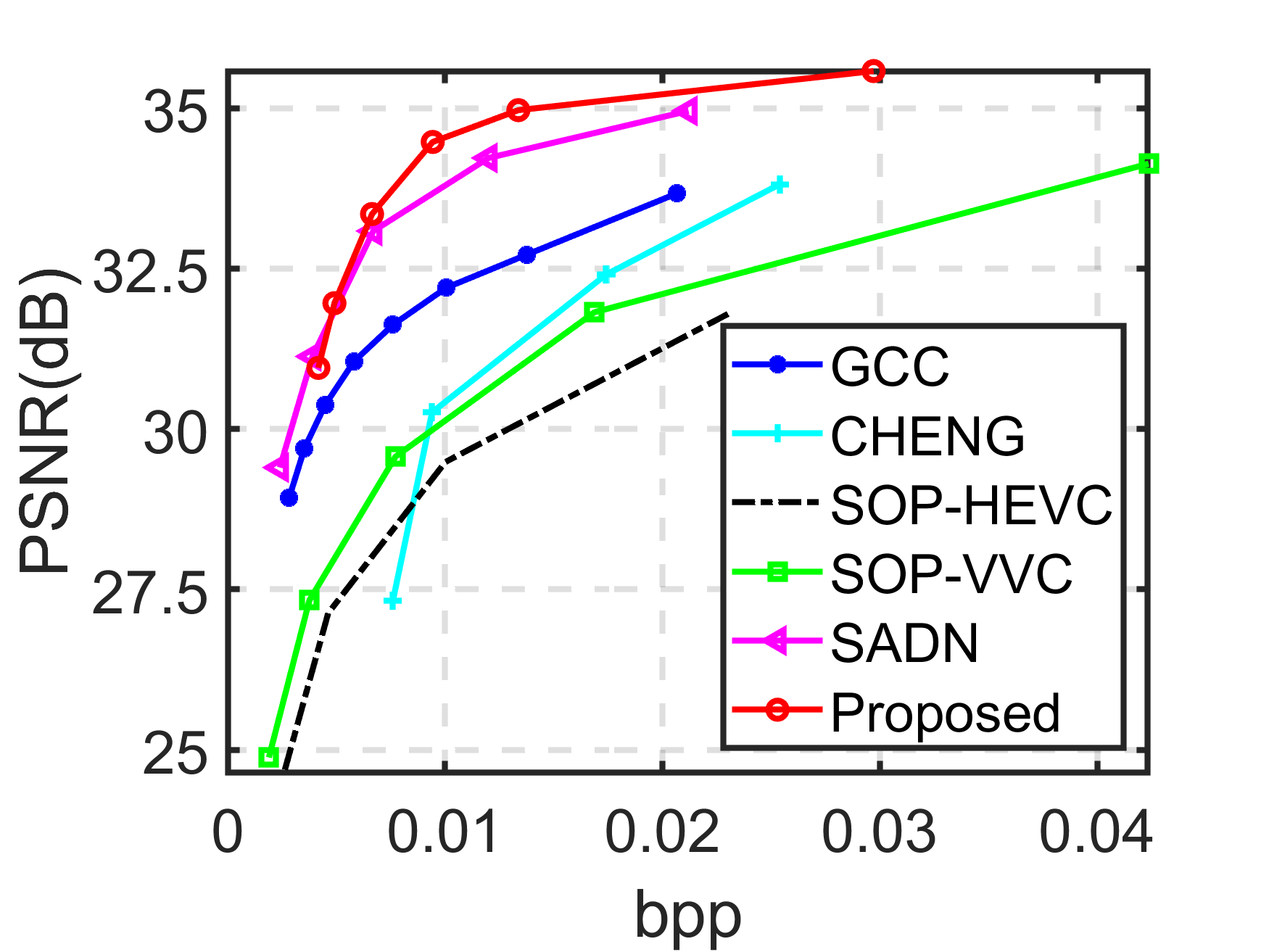}}
  \subfloat[Danger De Mort]{\includegraphics[width=0.25\linewidth]{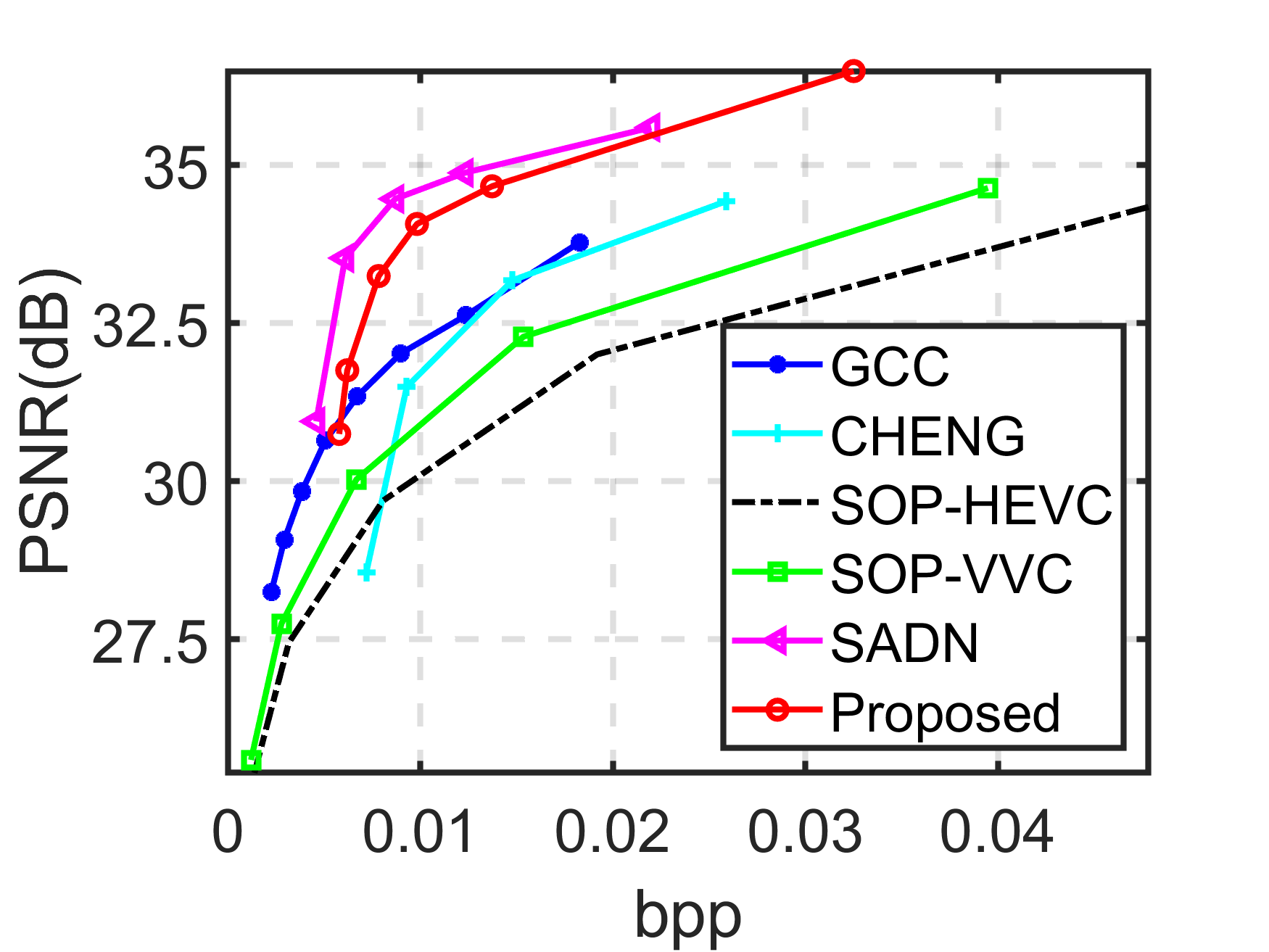}}
  \subfloat[Flowers]{\includegraphics[width=0.25\linewidth]{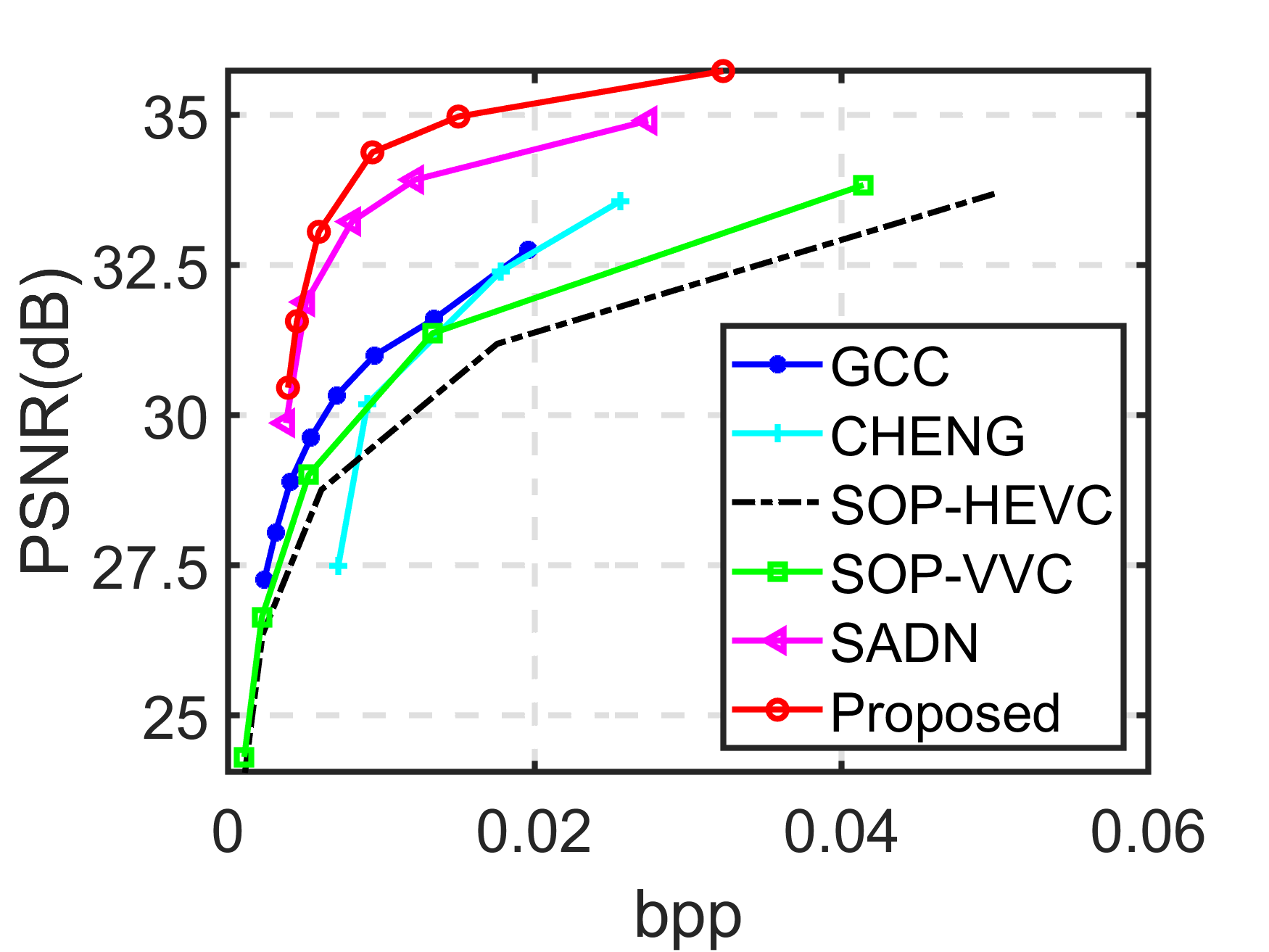}}
  \subfloat[Stone Pillars Outside]{\includegraphics[width=0.25\linewidth]{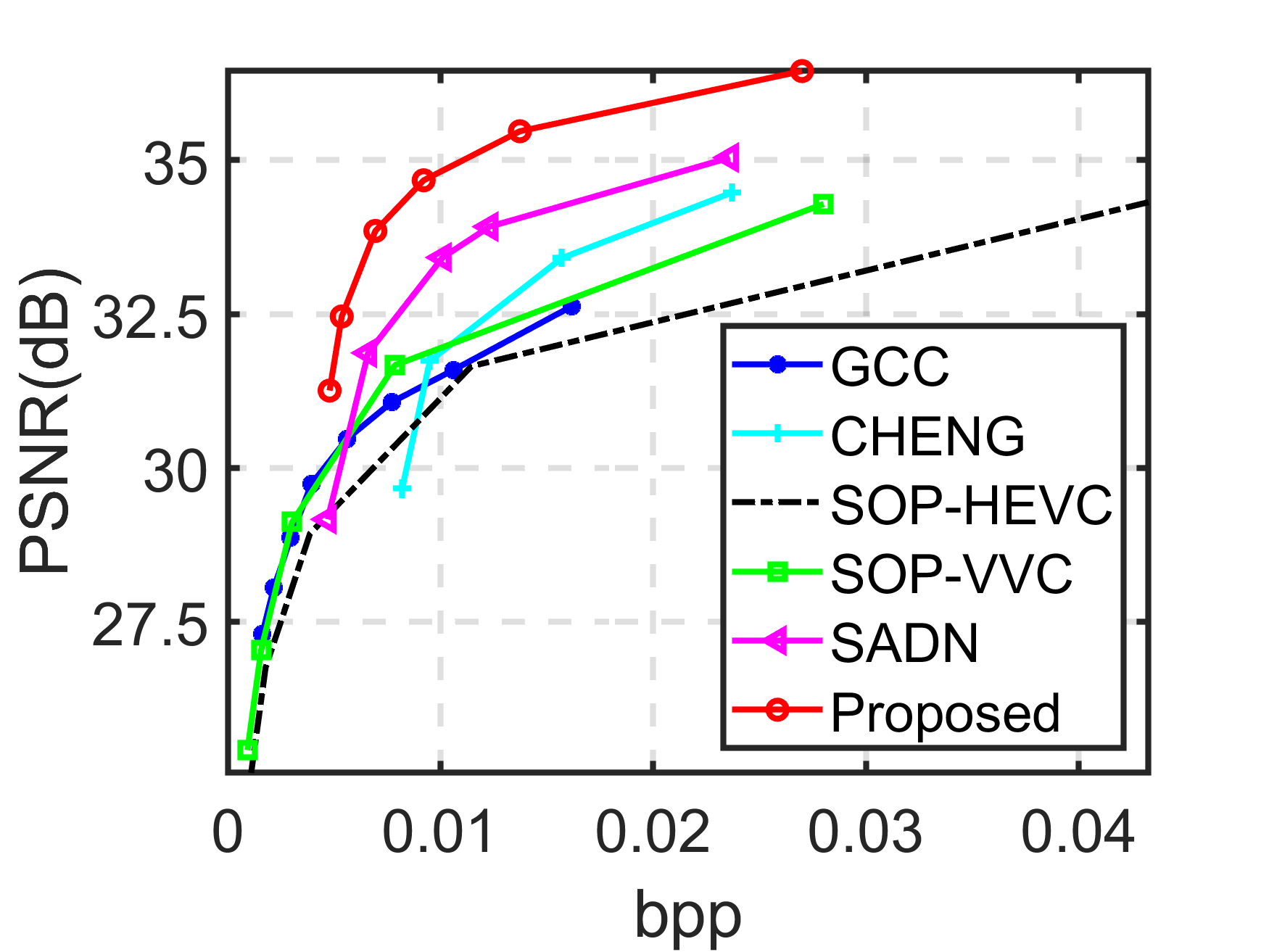}}\\
  \subfloat[Vespa]{\includegraphics[width=0.25\linewidth]{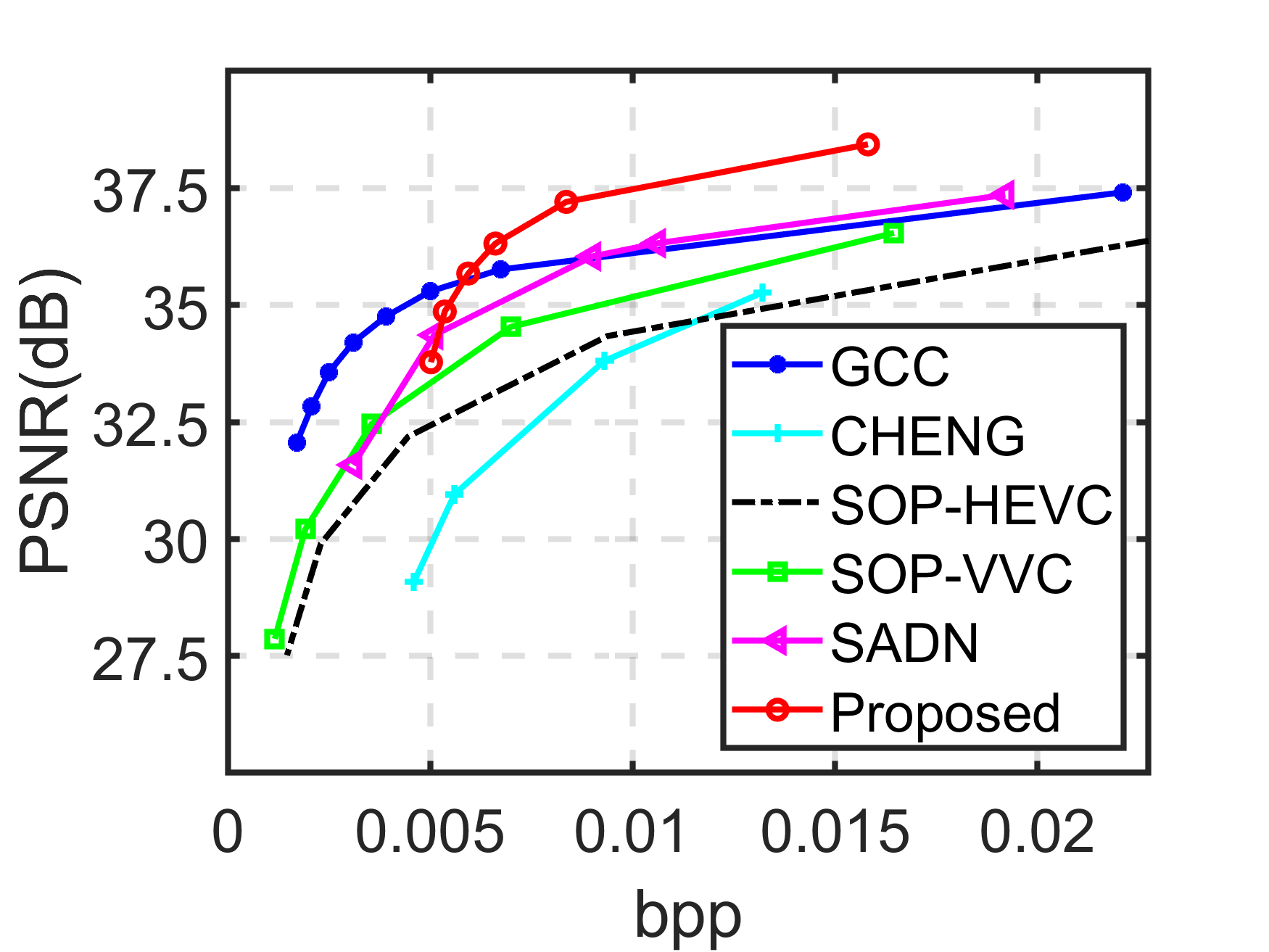}}
  \subfloat[Ankylosaurus\&Diplodocus]{\includegraphics[width=0.25\linewidth]{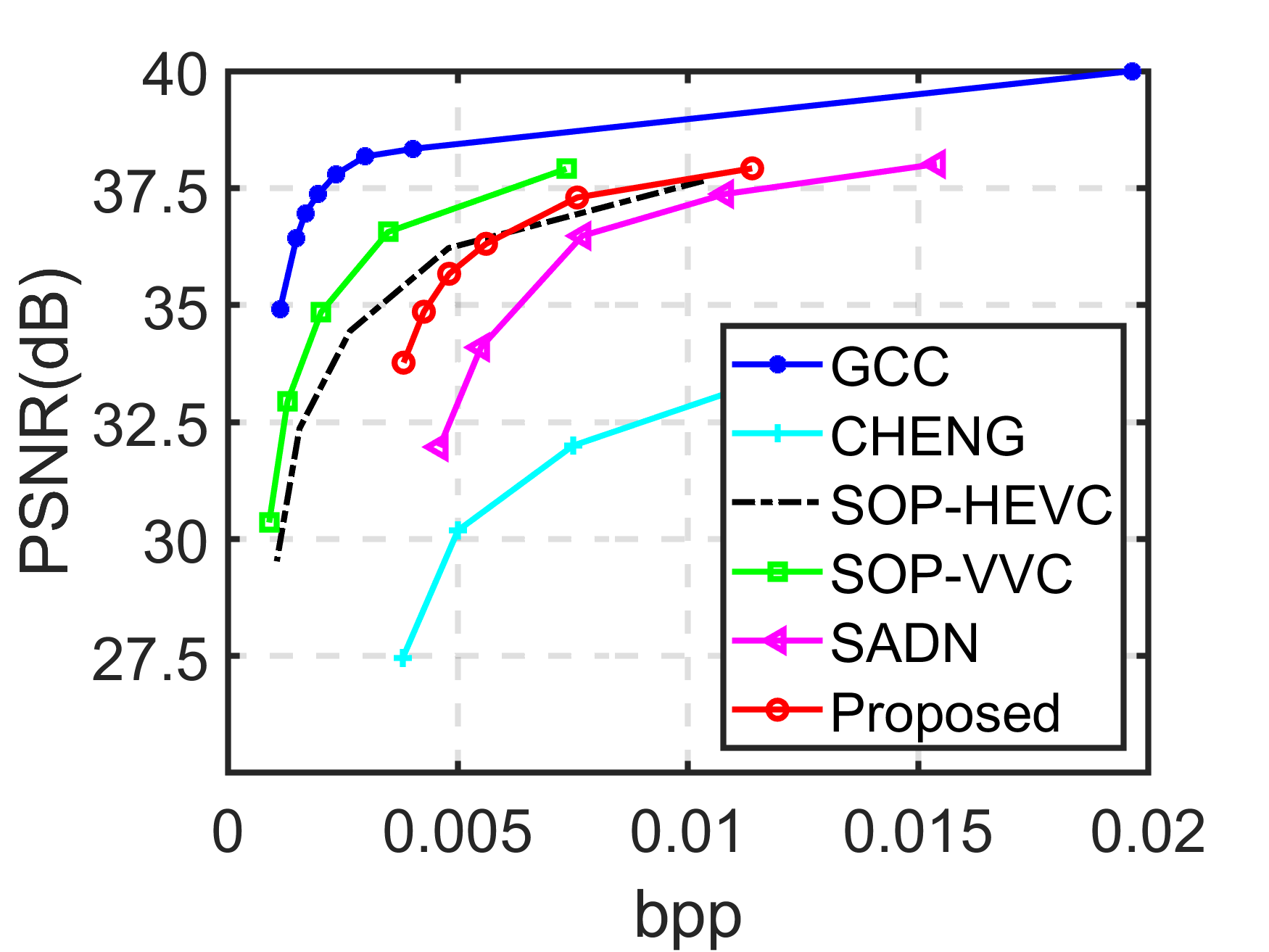}}
  \subfloat[Desktop]{\includegraphics[width=0.25\linewidth]{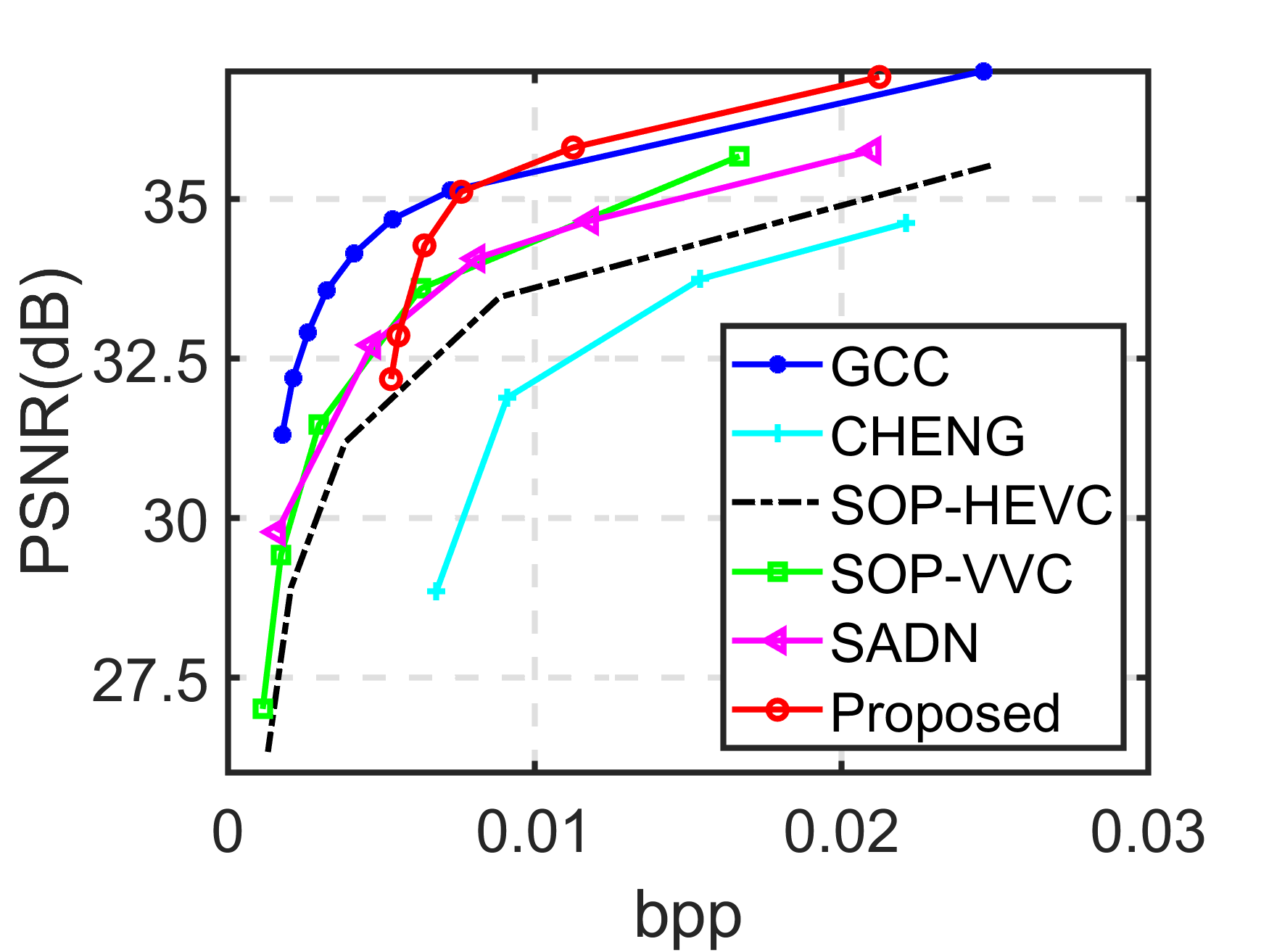}}
  \subfloat[Magnets]{\includegraphics[width=0.25\linewidth]{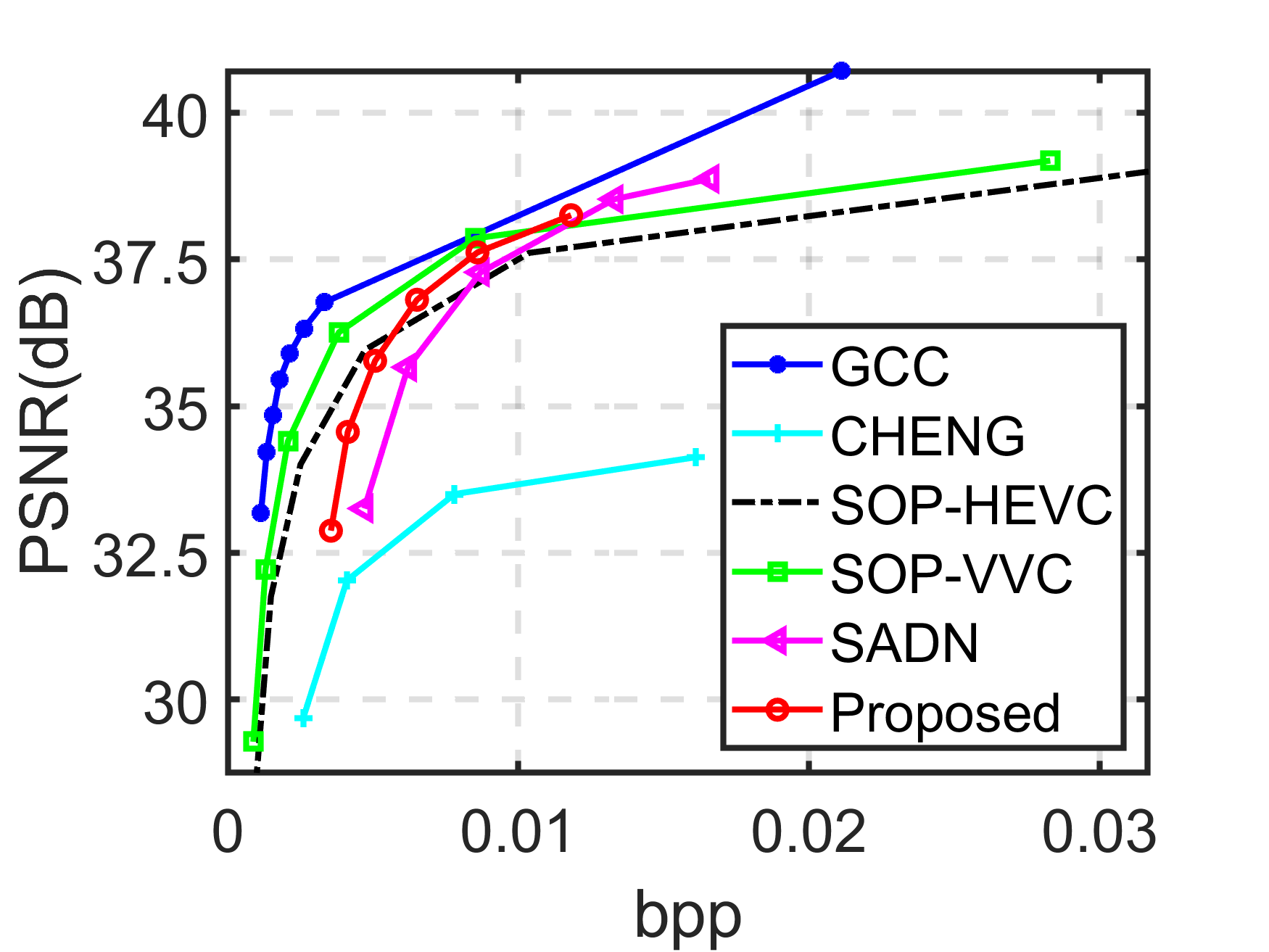}}\\
  \subfloat[Fountain\&Vincent]{\includegraphics[width=0.25\linewidth]{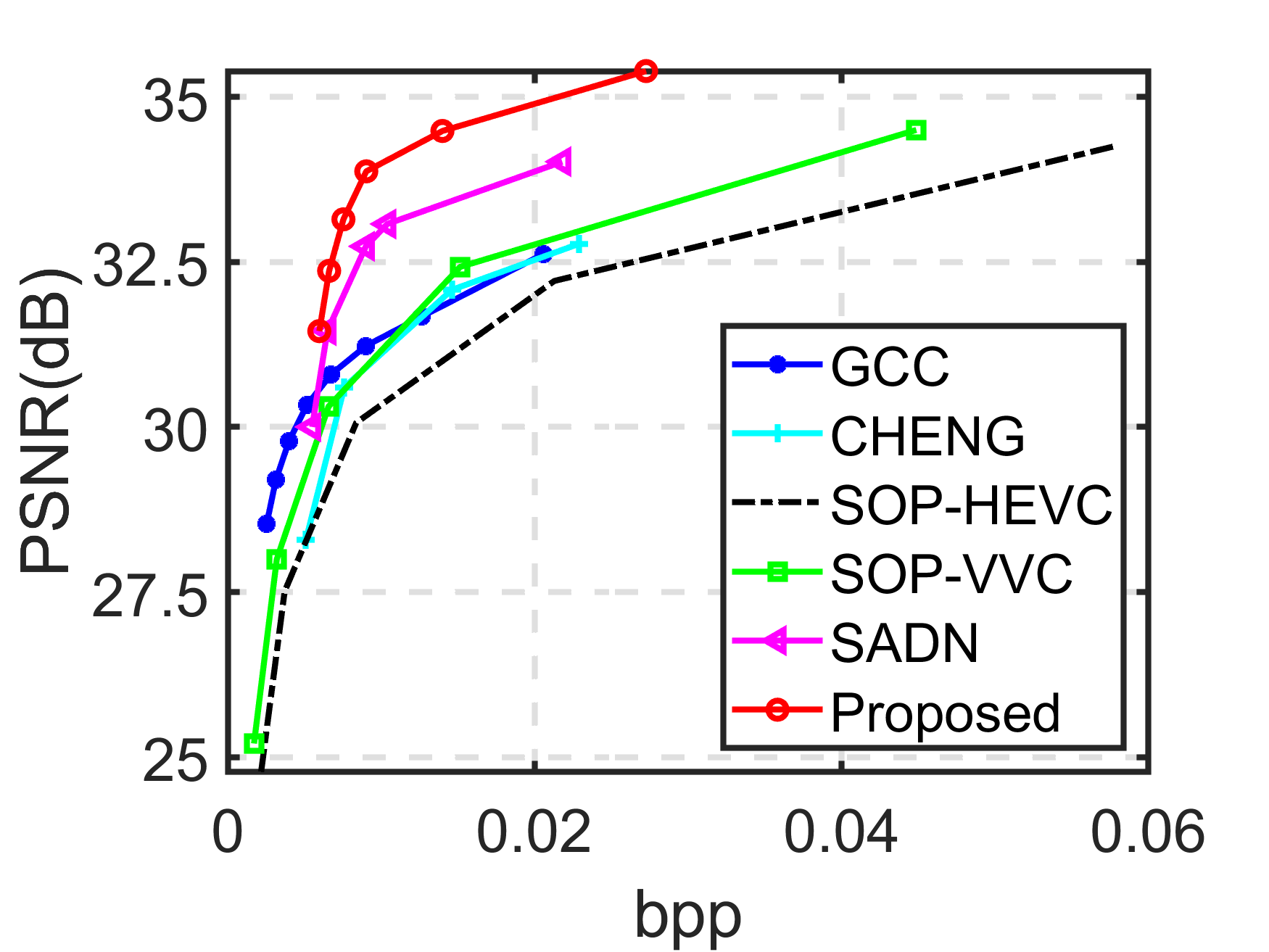}}
  \subfloat[Friends]{\includegraphics[width=0.25\linewidth]{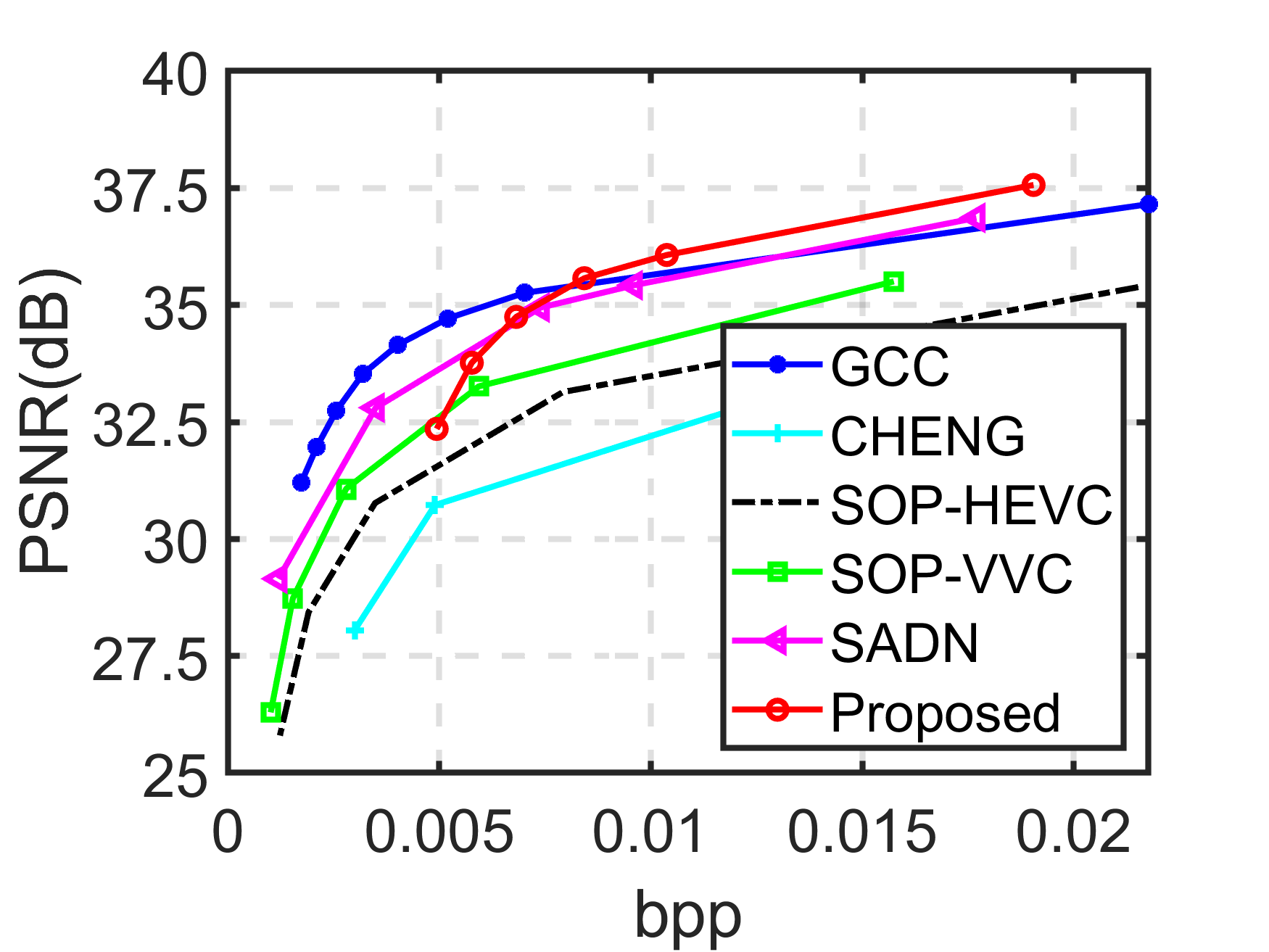}}
  \subfloat[Color Chart]{\includegraphics[width=0.25\linewidth]{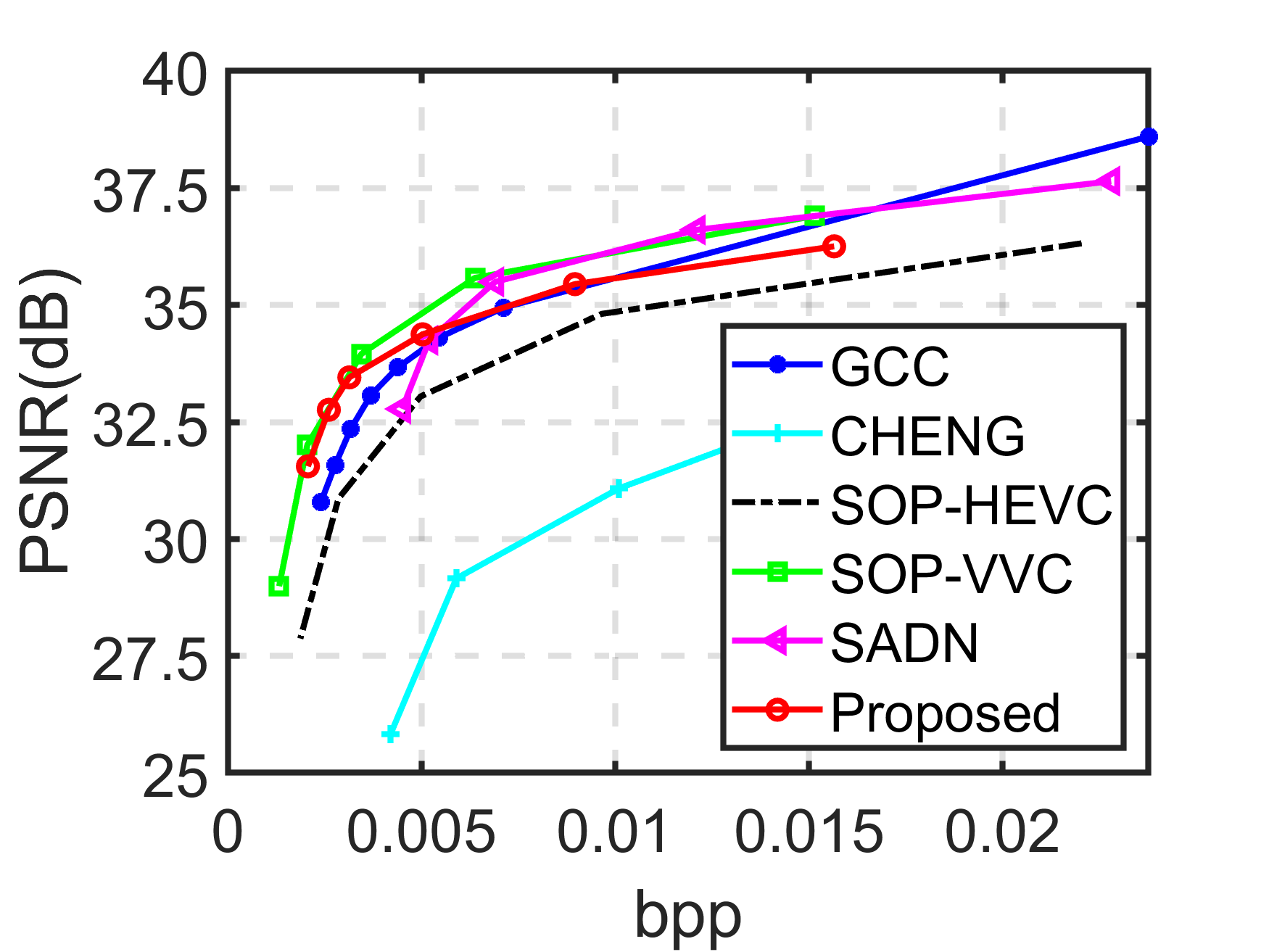}}
  \subfloat[ISO Chart]{\includegraphics[width=0.25\linewidth]{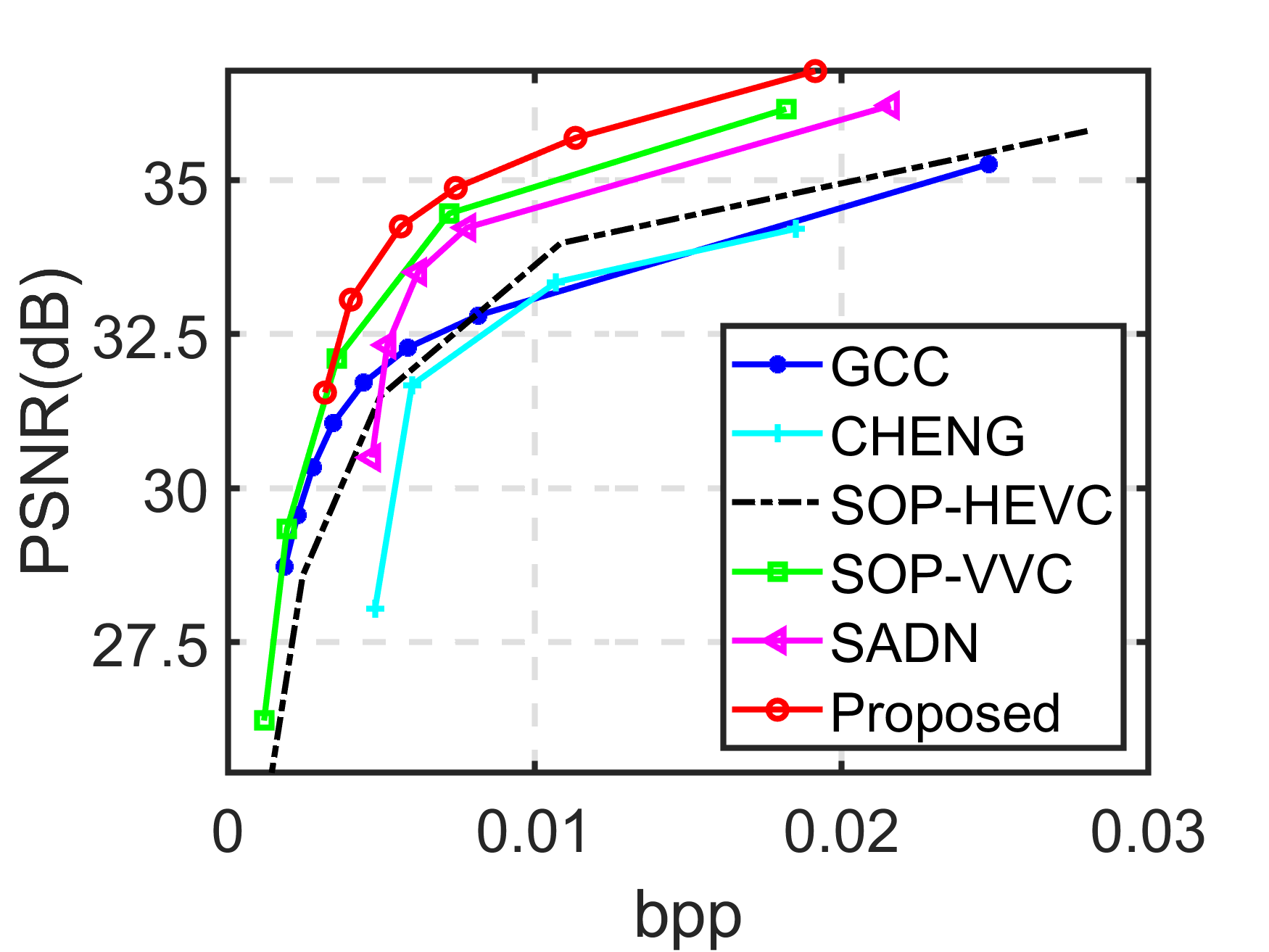}}
  \caption{LF coding performance comparison with the state-of-the-art works on the ICME Grand Challenge dataset.}
  \label{ddd}
\end{figure*}

\section{Experimental Results and Analysis}
\subsection{Experimental Settings}
We implemented the proposed LFIC-DRASC on the CompressAI platform \cite{CompressAI}. The dataset PINET proposed by \cite{SADN} was adopted for training, from which 40,000 MacPI based LF images were selected and randomly cropped with the size of 834$\times$834. All models were trained for 1.6M steps using the Adam optimizer \cite{Adam} with a batch size of 8. The whole architecture was trained on RTX 3090 GPU, and the CPU was Intel Core i9-10900X. The initial learning rate was $10^{-4}$, and reduced when the optimization stops improving. In specific, the learning rate scheduler function was the ReduceLROnPlateau in Pytorch2.0. When the model is optimized under MSE, the lambda values are set to 0.00015, 0.0002, 0.0006, 0.001, and 0.003 respectively. The number of potential and super-potential channels for our model is set to 64 and 16. Bj$\o$ntegaard metrics \cite{BD-BR}, including both Bj$\o$ntegaard Delta PSNR (BD-PSNR) and Bj$\o$ntegaard Delta bitrate (BD-BR), are adopted to measure coding performance.

\begin{figure*}[!htb]
  \begin{tabular*}{\textwidth}{@{\extracolsep{\fill}}r@{\hspace{0.0075\textwidth}} c c c }
  \raisebox{+0.1\textwidth}{Central SAI} & \includegraphics[width=0.28\textwidth]{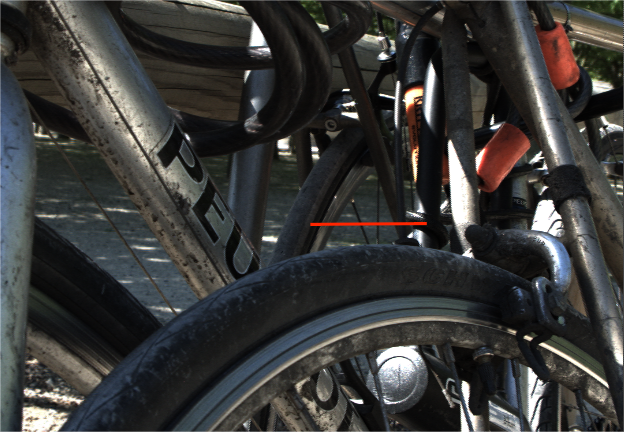} & \includegraphics[width=0.28\textwidth]{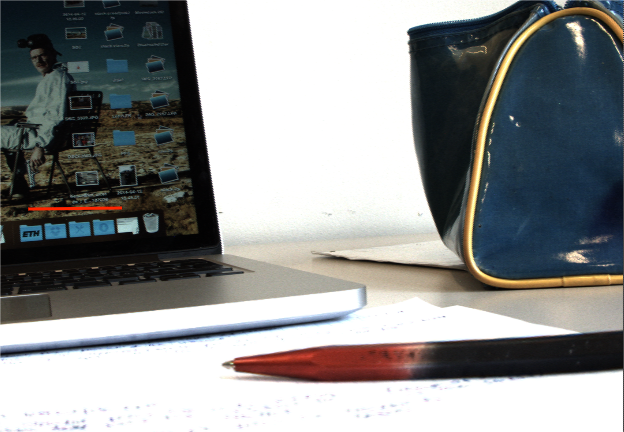} & \includegraphics[width=0.28\textwidth]{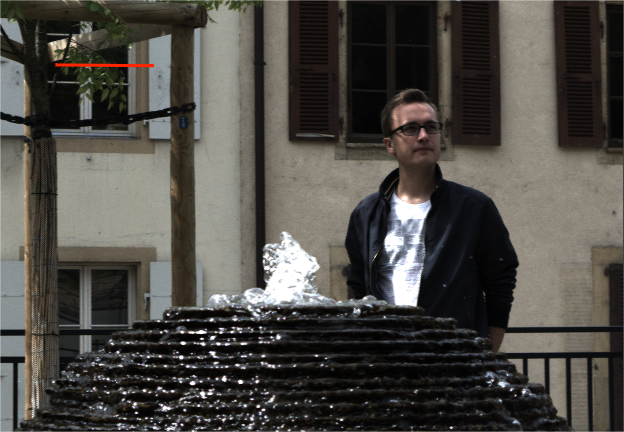} \\
  Original & \includegraphics[width=0.28\textwidth, height=0.07\textwidth]{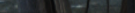} &\includegraphics[width=0.28\textwidth, height=0.07\textwidth]{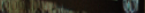} & \includegraphics[width=0.28\textwidth, height=0.07\textwidth]{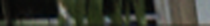}  \\
  & PSNR(dB)/bpp & PSNR(dB)/bpp & PSNR(dB)/bpp \\

  SADN & \includegraphics[width=0.28\textwidth, height=0.07\textwidth]{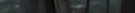} &\includegraphics[width=0.28\textwidth, height=0.07\textwidth]{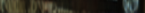}  & \includegraphics[width=0.28\textwidth, height=0.07\textwidth]{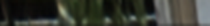} \\
   & 35.31/0.0203 & 35.76/0.0148 & 35.89/0.0148 \\
  GCC & \includegraphics[width=0.28\textwidth, height=0.07\textwidth]{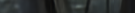}   &\includegraphics[width=0.28\textwidth, height=0.07\textwidth]{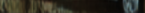}  & \includegraphics[width=0.28\textwidth, height=0.07\textwidth]{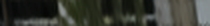}  \\
   & 33.29/0.0221 & 36.02/0.0169 & 34.45/0.0155 \\
  SOP-HEVC & \includegraphics[width=0.28\textwidth, height=0.07\textwidth]{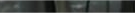}    &\includegraphics[width=0.28\textwidth, height=0.07\textwidth]{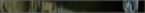}  & \includegraphics[width=0.28\textwidth, height=0.07\textwidth]{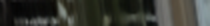} \\
   & 30.78/0.0225 & 33.52/0.0162 & 33.21/0.0212 \\
  SOP-VVC & \includegraphics[width=0.28\textwidth, height=0.07\textwidth]{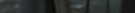}    &\includegraphics[width=0.28\textwidth, height=0.07\textwidth]{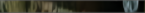}  & \includegraphics[width=0.28\textwidth, height=0.07\textwidth]{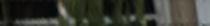} \\
   & 32.05/0.0214 & 34.87/0.0155 & 34.19/0.0174 \\

  Cheng's & \includegraphics[width=0.28\textwidth, height=0.07\textwidth]{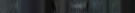}     &\includegraphics[width=0.28\textwidth, height=0.07\textwidth]{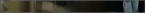}  & \includegraphics[width=0.28\textwidth, height=0.07\textwidth]{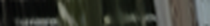}  \\
   & 31.41/0.0232 & 30.80/0.0141 & 32.77/0.0197 \\

  Proposed & \includegraphics[width=0.28\textwidth, height=0.07\textwidth]{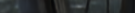}     &\includegraphics[width=0.28\textwidth, height=0.07\textwidth]{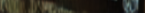} & \includegraphics[width=0.28\textwidth, height=0.07\textwidth]{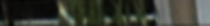}  \\
   & \textbf{35.74/0.0197} &  \textbf{36.42/0.0146} & \textbf{36.21/0.0140} \\

  \end{tabular*}
  \caption{Visual comparison results of EPI reconstructed using different codecs, where the red lines indicate selected EPI regions.}
  \label{fig:EPI}
  \end{figure*}


\subsection{Coding Performance Comparison}
We tested our model on the International Conference on Multimedia and Expo (ICME) 2016 Grand Challenge test dataset \cite{2016ICME}. The LF image thumbnails from this data set are shown in Fig. \ref{fig:all}. The proposed method was compared with the SOTA non-deep learning methods GCC\cite{GCCHuang2020LowBL}, SOP\cite{6-24}, and the learning based end-to-end LF image compression schemes of SADN\cite{SADN} and Cheng's\cite{ChengCVPR2020}.

As shown in Fig. \ref{ddd}, our proposed method achieves the best performance for most of LF images, which is remarkably higher than that of the other schemes. On the overall, the coding gains are significantly higher at high bit rates than at low bit rates. We find that if the image has a flat background and a uniformly sparse foreground (\emph{Ankylosaurus\&Diplodocus, Magnets, Color Chart}), our scheme is a little worse when compared to schemes of GCC, SOP-VVC. While traditional schemes handle uniform structures well, learning-based methods excel with complex textures (\emph{Bikes, Flowers, Stone Pillars Outside, and Fountain\&Vincent}), outperforming traditional compression. Compared with SADN, our scheme is superior in overall performance due to its further disentangle of LF and complete decomposition of complex 4D data. The Cheng's scheme, on the other hand, performs poorly on all LF images, which demonstrates that the LF data is more different from the traditional natural images, and our scheme is designed to be very effective for LF.
Table I presents BD-BR and BD-PSNR values, showing our method reduces bit rates by 59.8\%, 35.5\% and 20.5\% compared to SOP-VVC, SOP-HEVC and SADN, respectively. However, it does not always perform well, where the positive BD-BR values for \emph{Ankylosaurus\&Diplodocus, Magnets, Color Chart} are presented in this table, indicating further optimization for the proposed method is required.


\subsection{Visual Quality Comparison}

\begin{figure*}[!ht]
  \centering
  \setlength{\tabcolsep}{1pt}
  \captionsetup[subfloat]{font=footnotesize}

  \subfloat[Original (PSNR(dB)/bpp)]{\includegraphics[width=0.3\textwidth]{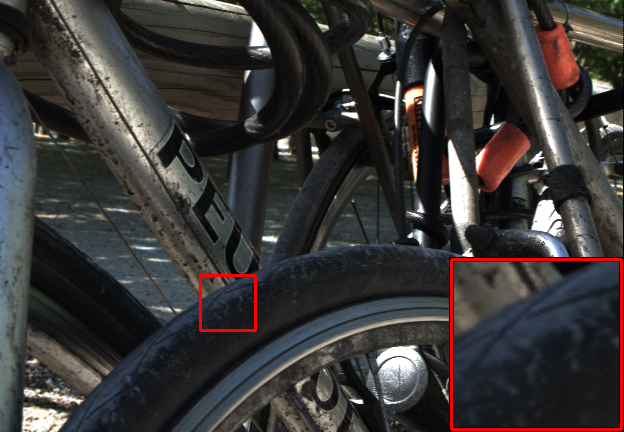}} \hfill
  \subfloat[Cheng's (34.41/0.0245)]{\includegraphics[width=0.3\textwidth]{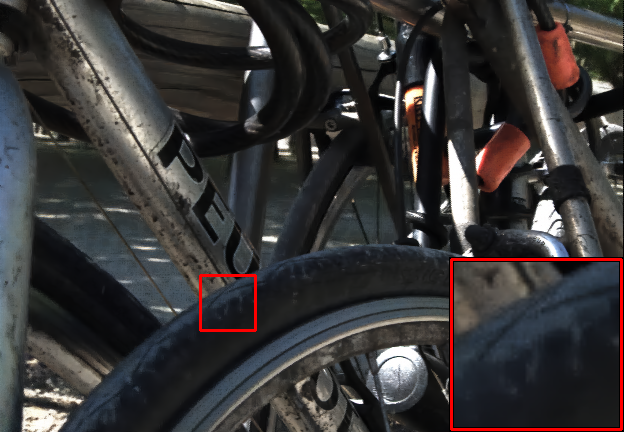}} \hfill
  \subfloat[SOP-HEVC (31.3/0.0233)]{\includegraphics[width=0.3\textwidth]{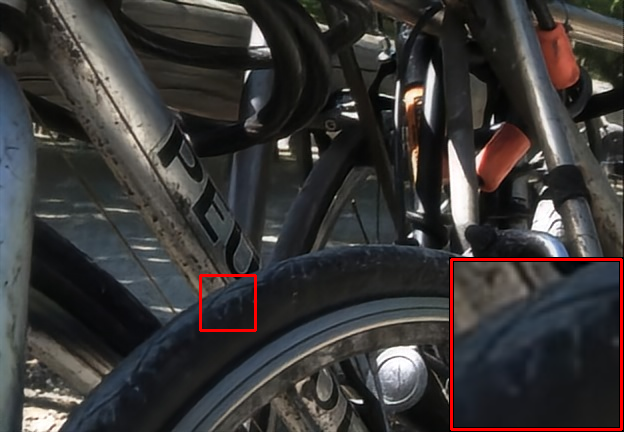}}\hfill
  \subfloat[GCC (37.14/0.0231)]{\includegraphics[width=0.3\textwidth]{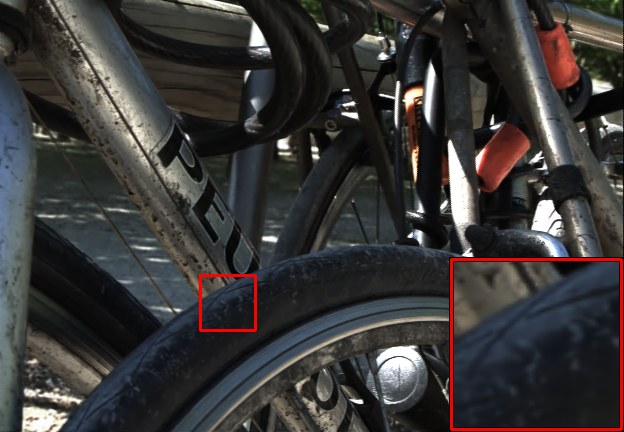}}  \hfill
  \subfloat[SADN (37.55/0.0225)]{\includegraphics[width=0.3\textwidth]{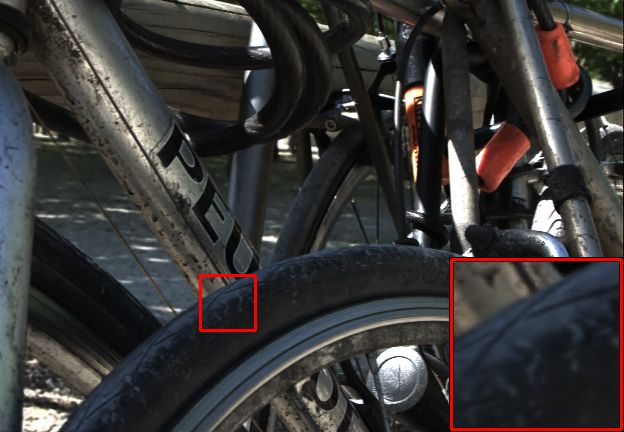}} \hfill
  \subfloat[Proposed (38.71/0.0227)]{\includegraphics[width=0.3\textwidth]{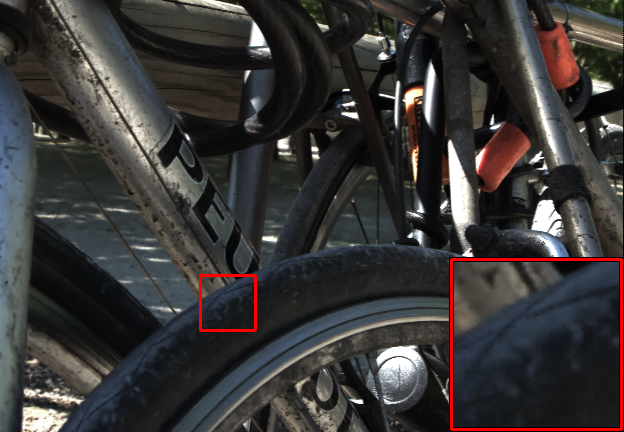}} \hfill

  \caption{Visual comparison results of central SAI from \emph{Bikes}, where PSNR and bpp values are presented.}
\label{fig:Views1}
\end{figure*}

\begin{figure*}[!ht]
  \centering
  \setlength{\tabcolsep}{1pt} 
  \captionsetup[subfloat]{font=footnotesize}
  \subfloat[Original (PSNR(dB)/bpp)]{\includegraphics[width=0.3\textwidth]{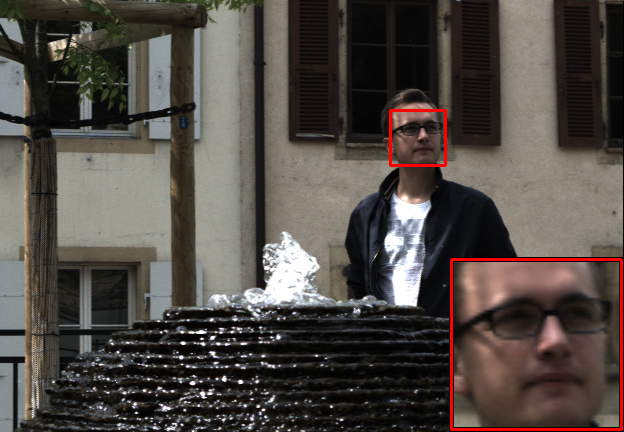}} \hfill
  \subfloat[Cheng's (32.44/0.0154)]{\includegraphics[width=0.3\textwidth]{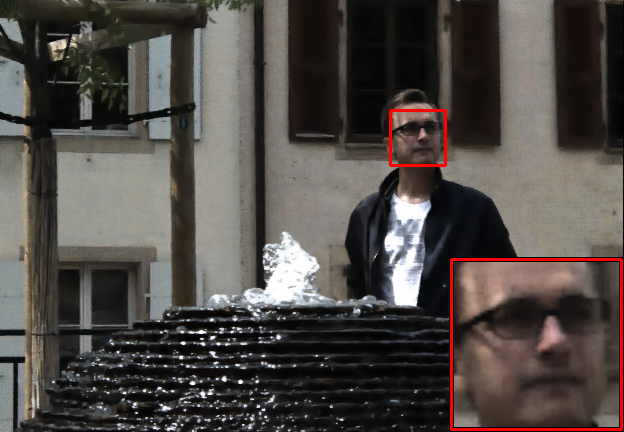}} \hfill
  \subfloat[SOP-HEVC (31.47/0.0142)]{\includegraphics[width=0.3\textwidth]{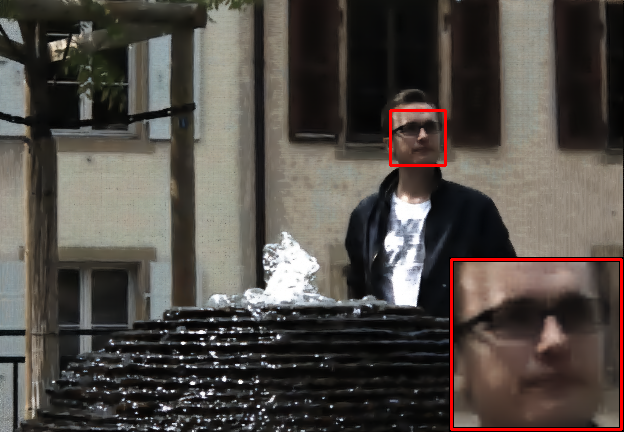}} \hfill
  \subfloat[GCC (33.92/0.0146)]{\includegraphics[width=0.3\textwidth]{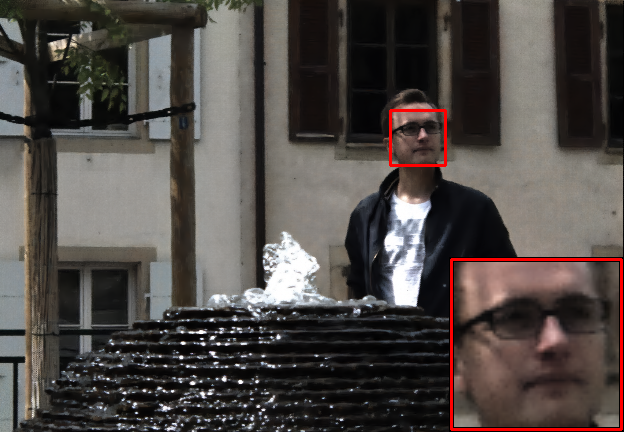}} \hfill
  \subfloat[SADN (35.79/0.0161)]{\includegraphics[width=0.3\textwidth]{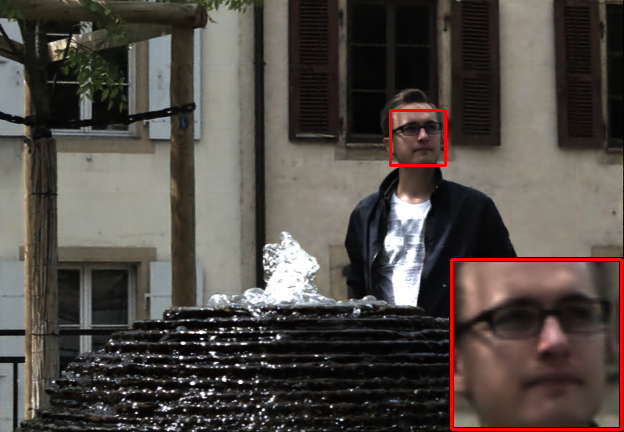}} \hfill
  \subfloat[Proposed (36.24/0.0155)]{\includegraphics[width=0.3\textwidth]{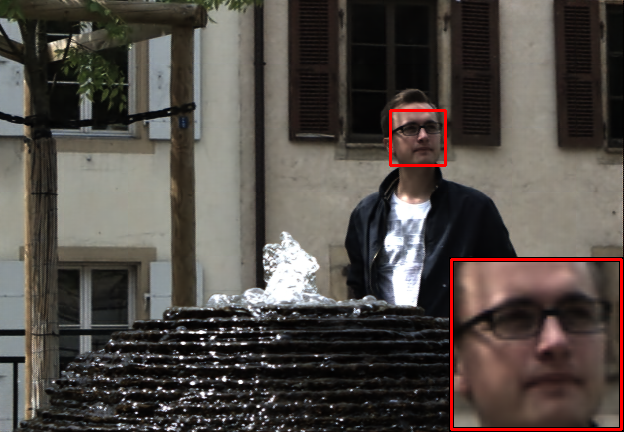}} \hfill
  \caption{Visual comparison results of central SAI from \emph{Fountain\&Vincent}, where PSNR and bpp values are presented.}
\label{fig:Views2}
\end{figure*}

Since the EPI contains the depth information of objects in the LF data, it can reflect the geometric consistency of the LF data. Three LF images and their EPIs were selected for visual comparison, \emph{i.e.}, \emph{Bikes}, \emph{Desktop}, and \emph{Fountain\&Vincent}. As shown in Fig. \ref{fig:EPI},
the two SOP-based methods, which convert the LF image into a PVS sequence for compression, result in distortion of the EPI due to accumulated errors. Cheng's method is less capable of modeling the details of the LF image and therefore suffers from large distortions. GCC obtains better performance than SOP-based and Cheng's methods because it maintains the consistency of the LF data. SADN demonstrates the effectiveness of the neural network model designed for the structure of the LF, showing better visual quality, but the disentangling of the LF is incomplete, and thus there is still improvement to be made in the reconstructed image. The proposed method adopts a more complete disentangling of the LF data, which is more capable of reconstructing LF than the SADN. The reconstructed image of our proposed method is more consistent with the linear representation of the original EPI image. This indicates that our proposed model better preserves the structure of LF and has a stronger ability to reconstruct LF images.

\begin{figure*}[!tb]
  \centering
  \subfloat[]{
    \includegraphics[height=0.25\textwidth]{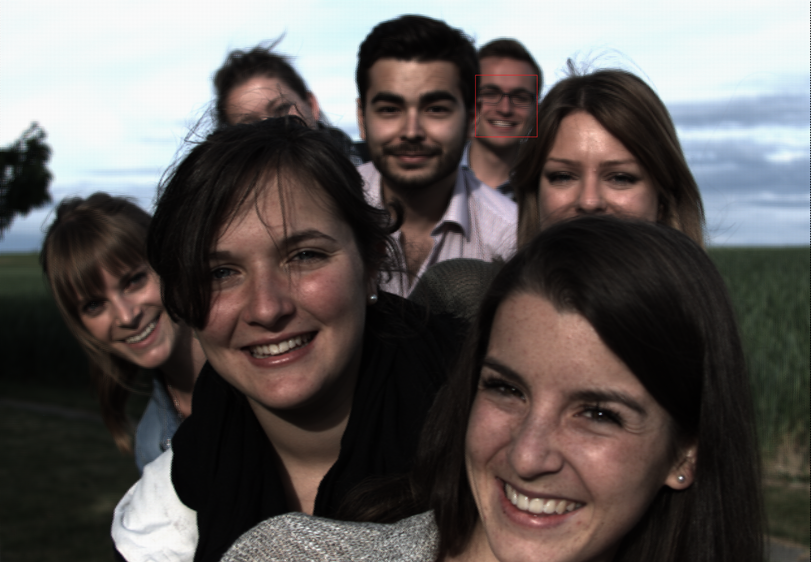}}
  \hspace{0.05cm}
  \subfloat[]{
    \includegraphics[height=0.25\textwidth]{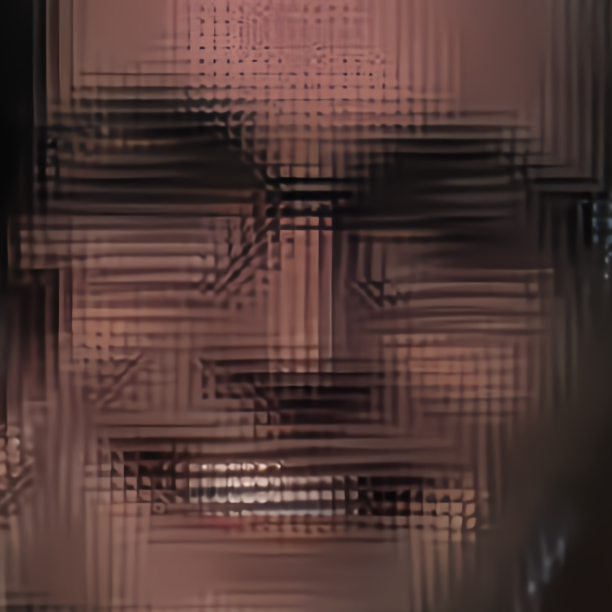}} 
      \hspace{0.05cm}
  \subfloat[]{
    \includegraphics[height=0.25\textwidth]{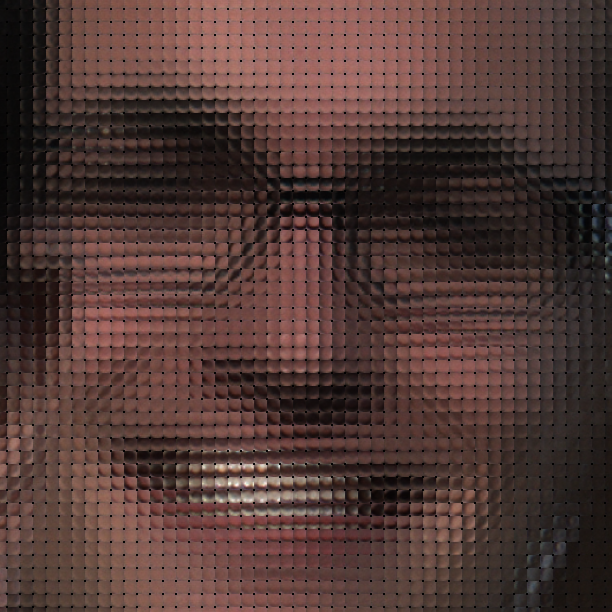}} 
      \hspace{0.05cm}
  \caption{Visual compression of MacPI.(a) Friends image, (b) Cheng's (34.07/0.0185), (c) Proposed (37.13/0.0174)}
  \label{fig:face}
\end{figure*}

Since both MacPI and EPI are not directly visualized by the human, we show the central SAI images for visual comparison, as shown in Figs. \Ref{fig:Views1} and \Ref{fig:Views2}. It is found that our scheme has more pleasing reconstruction details and produces a much less blurred view. Finally, we illustrate the visual differences from LF encoding and natural image encoding, as shown in Fig. \ref{fig:face}. It is found the proposed LFIC-DRASC can model the texture relationships between macro-pixels in MacPI and can capture the boundaries within macro-pixels through strip convolution, which enables a better image reconstruction.

\subsection{Ablation Study}

\begin{figure}
  \center
  \includegraphics[width=0.47\textwidth]{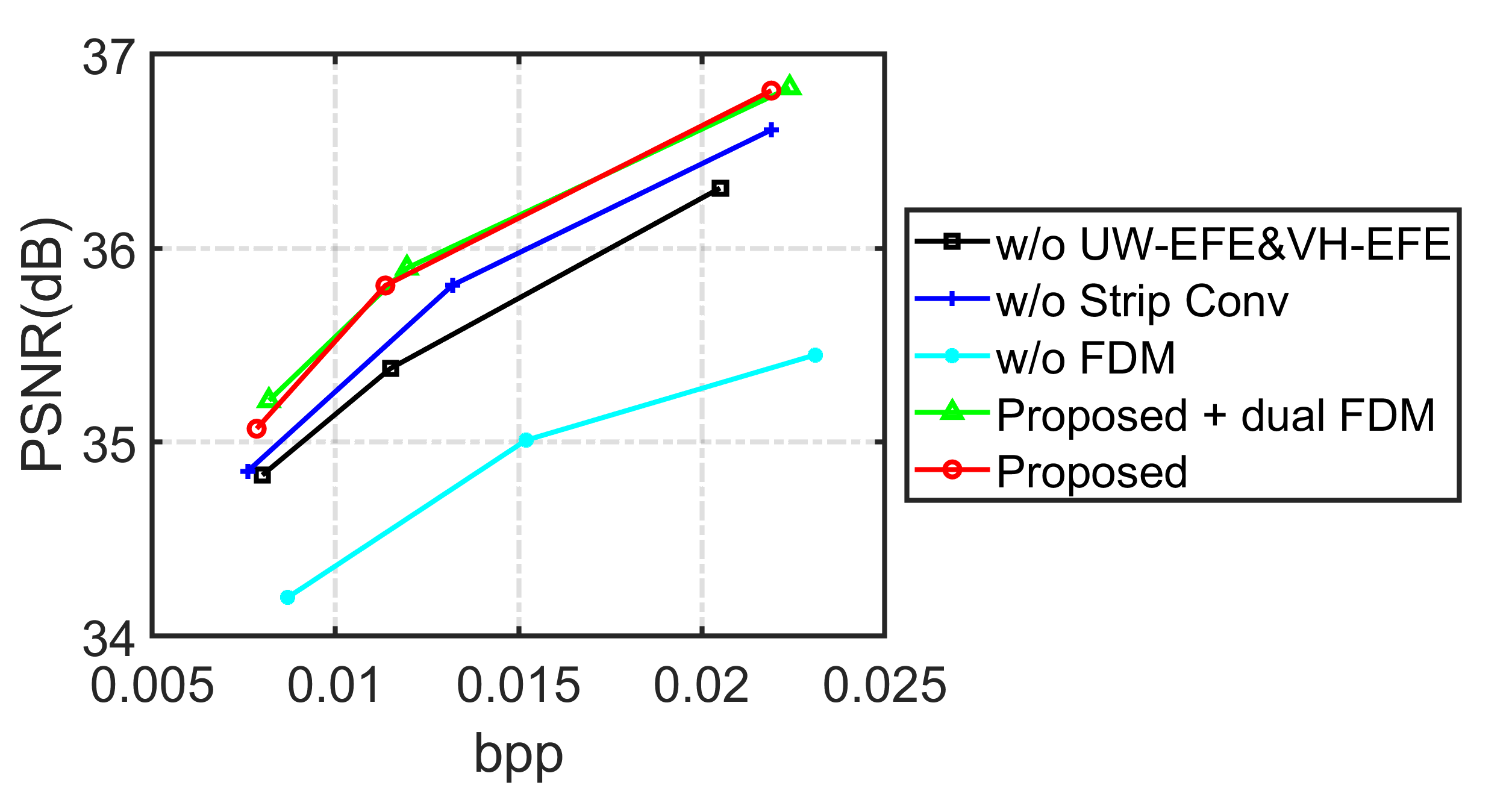}
  \caption{Ablation studies on the ICME Grand Challenge dataset.}
  \label{fig:ablation}
  \end{figure}

To validate the effectiveness of our proposed FDM module and asymmetric strip convolution, we conducted ablation experiments. All networks were trained using the same training parameters to assess the contribution of each module. There are five models in total, including: (1) the proposed model, (2) the proposed model without strip convolution, (3) the proposed model without FDM (referred to as w/o FDM), (4) the proposed model with added dual FDM, and (5) the proposed model without UW-EFE and VH-EFE feature extractors. As shown in Fig. \ref{fig:ablation}, the FDM module provides the greatest coding gain in our proposed method. This is because, for simple CNNs, extracting information from LF images without the aid of low-dimensional projections is challenging, and the entangled four-dimensional LF information is difficult for the network to model. This suggests that disentangling LF images in end-to-end LF compression is highly beneficial. We further explored the feasibility of adding FDM at the decoding side, specifically at the module of $g_s$. Considering that in end-to-end encoding, a ``reverse" feature extraction module is often added to ensure symmetry, we found that adding dual FDM did not significantly improve performance but increased encoding and decoding time. Therefore, for higher coding performance, we decided to only retain the FDM at the encoding side, preserving its original feature extraction physical meaning without significantly adding reverse FDM. Meanwhile, we further explored the proposed operators, namely UW-EFE and VH-EFE, to more thoroughly disentangle the LF data, allowing it to be projected into six different subspaces. The neural network can more easily model the LF in different subspaces, thus achieving greater coding gain. Finally, our proposed asymmetric strip convolution can establish long-term dependencies of LF data better than simple 3$\times$3 convolution due to the inherent long-range dependencies of LF data in both horizontal and vertical directions. Therefore, after adding ASC to the network, significant coding gain is achieved.

\begin{table}[t]\caption{Computational complexity comparison.} \label{table1}
\begin{center}
    \begin{tabular}{c|c|c|c|c}
    \cline {1-5} {Method}&{En. Time}&{De. Time}&{Params.}&{MACs}\\
    \hline
    {Cheng's}   &4314ms&5314ms&13.18M&180.25G\\
    {SADN}      &5139ms&6857ms&5.41M&206.97G\\
    {Proposed}  &5584ms&8721ms&3.72M&265.72G\\
    {Proposed/dual FDM} &5767ms&12455ms&5.65M&340.25G\\
    \hline
   \end{tabular}
\end{center}
\end{table}

\subsection{Computational Complexity}
In addition, we compared the computational complexity of the proposed method with two end-to-end image compression methods (\emph{i.e.}, Cheng's method and SADN) as well as the scheme with added dual FDM. The values of encoding time, decoding time, network parameters, and MACs are presented in Table \ref{table1}. It can be observed that after adding the dual FDM, the decoding time and MACs increased significantly. Considering overall performance, removing the FDM at the decoding side brought the encoding and decoding time to an acceptable range. Moreover, since both our scheme and SADN use a relatively low channel count of 48, instead of the 256 channels commonly used in end-to-end image encoding like Cheng's, the overall parameters are fewer. Overall, the computational complexity of the proposed method remains a challenging issue, with encoding time longer than Cheng's method and SADN. Optimization of computational complexity is expected to be addressed in future work.

\section{Conclusions}
In this paper, we propose a deep Light Field (LF) Image Compression using Disentangled Representations and Asymmetrical Strip Convolution (LFIC-DRASC). Firstly, we formulate the LF compression as a joint problem of LF feature representation and image compression. Secondly, to represent the LF feature more effectively, we propose two novel LF feature extractors and a Feature Disentangling Model (FDM). Thirdly, we propose the LFIC-DRASC network for LF image compression, which consists of FDM and Strip Convolution Module (SCM) based variational autoencoder. In the SCM, two Asymmetrical Strip Convolution (ASC) operators, i.e. horizontal and vertical ones, are proposed to capture long-range correlation in LF feature space. Experimental results show that our proposed LFIC-DRASC can effectively achieve a higher compression ratio and better visual quality.

\bibliographystyle{IEEEtran}
\bibliography{IEEEabrv, reference}

\end{document}